\documentclass[prl,aps,twocolumn,longbibliography]{revtex4-1}

\pdfoutput=1

\date{\today}

\newcommand{\kB}{k_{\textrm{B}}}

\newcommand{\ts}{t_{\textrm{s}}}
\newcommand{\Us}{U_{\textrm{s}}}

\usepackage{float}

\usepackage{fourier}

\usepackage{color}
\usepackage{graphicx}
\usepackage{amsmath,amssymb}

\usepackage{times}
\usepackage{upgreek}
\usepackage{psfrag} 
\usepackage{latexsym} 
\usepackage{amstext}
\usepackage{amsxtra} 

\usepackage{textcomp}
\usepackage{amsfonts}
\usepackage{graphicx}
\usepackage{bm}
\usepackage{color}
\usepackage{braket}
\usepackage{units}
\usepackage[svgnames]{xcolor}

\definecolor{myColor}{rgb}{0.02,0.12,0.3}
\definecolor{myciteColor}{rgb}{0.39,0.7,0.89}
\usepackage[colorlinks=true,citecolor=myColor,linkcolor=myColor,urlcolor=myColor]{hyperref}

\def\be{\begin{equation}}
\def\ee{\end{equation}}

\begin{document} 

\title{
Many-body Decay of the Gapped Lowest Excitation of a Bose--Einstein Condensate\\
}

\author{
Jinyi Zhang,$^{1,{\color{myColor}\ast}}$
Christoph Eigen,$^{1,{\color{myColor}\ast},{\color{myColor}\dag}}$
Wei Zheng,$^{1,2,3,{\color{myColor}\dag}}$
Jake A. P. Glidden,$^{1}$
Timon A. Hilker,$^{1}$
Samuel J. Garratt,$^{1,4}$
Raphael Lopes,$^{1,5}$
Nigel R. Cooper,$^{1}$
Zoran Hadzibabic,$^{1}$ and
Nir Navon$^{6}$
}

\affiliation{
\vspace{2mm}$^1$ Cavendish Laboratory, University of Cambridge, J. J. Thomson Avenue, Cambridge CB3 0HE, United Kingdom\\
$^2$ Hefei National Laboratory for Physical Sciences at the Microscale and Department of Modern Physics, University of Science and Technology of China, Hefei 230026, China\\
$^3$ CAS Center for Excellence in Quantum Information and Quantum Physics, University of Science and Technology of China, Hefei 230026, China\\
$^4$ Theoretical Physics, University of Oxford, Parks Road, Oxford OX1 3PU, United Kingdom\\
$^5$ Laboratoire Kastler Brossel, Coll{\`e}ge de France, CNRS, ENS-PSL University, UPMC-Sorbonne Universit{\'e}s, 11 Place Marcelin Berthelot, F-75005 Paris, France\\
$^6$ Department of Physics, Yale University, New Haven, Connecticut 06520, USA
}

\begin{abstract}	
We study the decay mechanism of the gapped lowest-lying excitation of a quasi-pure box-trapped atomic Bose--Einstein condensate. Owing to the absence of lower-energy modes, or direct coupling to an external bath, this excitation is protected against one-body (linear) decay and the damping mechanism is exclusively nonlinear.
We develop a universal theoretical model that explains this fundamental nonlinear damping as a process whereby two quanta of the gapped lowest excitation mode couple to a higher-energy mode, which subsequently decays into a continuum.
We find quantitative agreement between our experiments and the predictions of this model. 
Finally, by strongly driving the system below its (lowest) resonant frequency we observe third-harmonic generation, a hallmark of nonlinear behavior.

\end{abstract}

\maketitle 

Our understanding of quantum many-body systems is rooted in the existence of elementary excitations such as quasi-particles and phonons.  
The nature of these excitations reflects the form of order in the underlying state of matter. Moreover, low-temperature thermodynamics and near-equilibrium transport properties can be computed from the energy spectrum of these modes.

Typically this energy spectrum can be calculated to lowest order by assuming that the excitations are non-interacting, and thus have an infinite lifetime. Taking into account the interactions between the quasi-particles in the continuum limit generically leads to finite lifetimes, even at zero temperature~\cite{fetter2012quantum}.  

On the other hand, the lifetime of gapped excitations has been far less investigated. Gaps in the energy spectra naturally arise in finite-size systems \cite{henley1964energy,ashoori1996electrons} or from many-body effects in infinite systems \cite{cooper1956bound,kohn1964theory}, or a combination of the two \cite{garcia2008bardeen}.
In either case the lifetime of the lowest-lying excitation of a many-body system at zero temperature is of particular interest since it is energetically immune to any one-body (\emph{i.e.}~linear) decay mechanism. 
In this Letter, we study this fundamental many-body problem experimentally and theoretically by investigating the decay of the lowest excitation of a homogeneous box-trapped atomic Bose--Einstein condensate (BEC). 

\begin{figure}[b!]
	\centerline{\includegraphics[width=1\columnwidth]{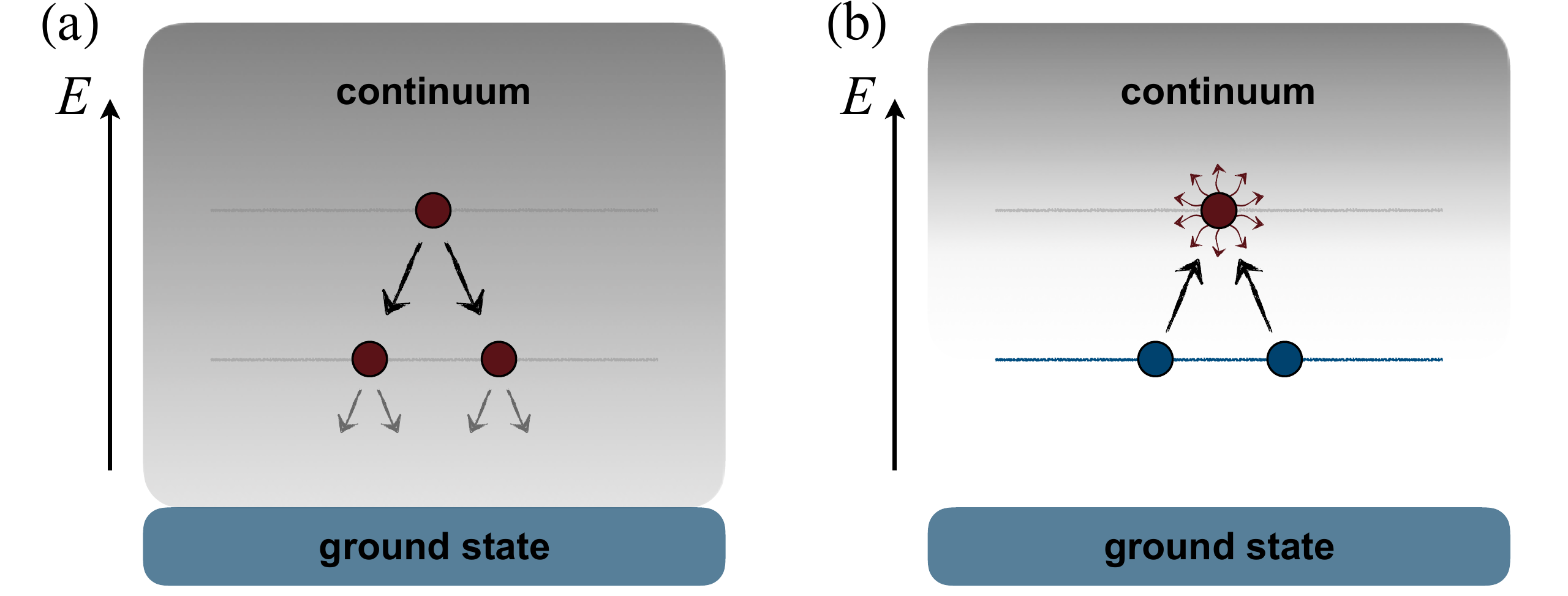}}
	\caption{Decay mechanisms of elementary excitations in a quantum many-body system. (a) Beliaev-like linear damping of an excitation in the continuum to two (or more) lower-lying excitations. (b) Nonlinear damping of the lowest-energy, gapped excitation (blue) to a higher-lying mode (red). The continuum is represented as a gray shade, while the horizontal lines indicate relevant energy levels.} 
	\label{Fig1}
\end{figure}

The weakly-interacting bulk Bose gas has been a remarkable testbed for exploring the physics of excitations and their interactions~\cite{jin1996collective,mewes1996collective,andrews1997propagation,stamper1998collisionless,stamper1999excitation,stenger1999bragg,marago2000observation,aboshaeer2001observation,steinhauer2002excitation,vogels2002experimental,katz2002beliaev,khaykovich2002formation,strecker2002formation,coddington2003observation,bretin2003quadrupole,engels2007observation,papp2008bragg,meppelink2009sound,jaskula2012acoustic,lopes2017quasiparticle,clark2017collective,ville2018sound,garratt2019from,feng2019correlations}. At low temperature, the excitations of an infinite uniform BEC have an energy given by the Bogoliubov spectrum; in the long-wavelength limit these excitations are phonons~\cite{bogoliubov1947on}. The leading decay channel for the phonons is a linear process, in which they spontaneously break up into pairs of longer-wavelength phonons [Fig.~\ref{Fig1}(a)], a damping mechanism first predicted by Beliaev~\cite{beliaev1958energy}. 

By contrast, in our case the finite system size leads to experimentally resolvable gaps in the excitation spectrum, and recent works suggested that the damping of the lowest mode is fully nonlinear (within experimental precision)~\cite{navon2016emergence, garratt2019from}.
Here we experimentally and theoretically elucidate the damping mechanism of this mode; we show that it can be explained by a generic model based on an inverse Beliaev-like process, whereby two elementary excitations merge into a higher-energy one [Fig.~\ref{Fig1}(b)]. 

%%%%%%%%%%%%%%%%%%%%%%%%%%%%%%%%%%%%%%%%%%
%%%%%%%%%%%%%%%%%%%%%%%%%%%%%%%%%%%%%%%%%%% 

%%%%%%%%%%%%%%%%%%%%%%%%%%%%%%%%%%%%%%%%%%
\begin{figure*}[t]
	\centerline{\includegraphics[width=\textwidth]{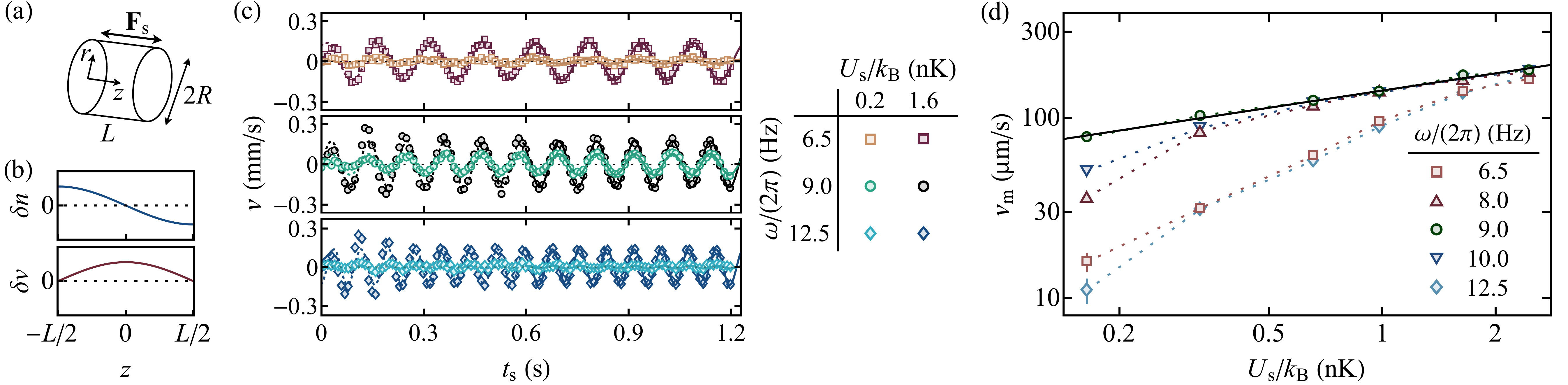}}
	\caption{Nonlinear dynamics of a driven box-trapped Bose--Einstein condensate.
	(a) Sketch of the optical box trap, highlighting the axis along which the drive ${\bf F}_\text{s}(t)$ is applied. (b) Sketch of the fundamental mode density $\delta n(z)$ and velocity $\delta v(z)$ profiles along the box axis.
	(c) Center-of-mass velocity response $v(t_\text{s})$ for three different drive frequencies $\omega$, close to the resonant $\omega_d/(2\pi) = 9.0(1)$~Hz, and two different drive strengths $\Us$ (see legend).
	(d) Steady-state response amplitude $v_\textrm{m}$ versus $\Us$ on log-log scale, for various drive frequencies (see legend). The on-resonance $v_\textrm{m}$ exhibits power-law behavior, and a fit to the data (solid line) gives $v_\textrm{m}\propto \Us^{0.32(1)}$. For progressively larger $\Us$ a wider range of $\omega$ become effectively resonant. The dotted lines are guides to the eye, and the error bars represent fitting uncertainties.
	}
	\label{Fig2}
\end{figure*}
%%%%%%%%%%%%%%%%%%%%%%%%%%%%%%%%%%%%%%%%%%% 
 
Our experiment starts with a quasi-uniform BEC of $^{87}$Rb atoms~\cite{gaunt2013bose}, prepared in a cylindrical optical box trap of radius $R=16(1)~\upmu$m and length $L=26(1)~\upmu$m [see Fig.~\ref{Fig2}(a)].
Our condensates consist of $N= 1.2(1)\times 10^5$ atoms, their chemical potential is $\mu\approx \kB \times 2.1$~nK, their healing length $\xi\approx1.1~\upmu$m, and the excitation frequency of the lowest-lying axial mode [Fig.~\ref{Fig2}(b)] is $\omega_d = 2\pi \times 9.0(1)$~Hz (see \cite{garratt2019from} for details). We excite this mode by applying a spatially uniform oscillating force ${\bf F}_\text{s}({\bf r},t)=(\Us/L)  \sin(\omega t) {\bf e}_z$, where ${\bf e}_z$ is the unit vector along the symmetry axis of the cylinder and $\Us$ is the maximal potential-energy drop across the box. Following a variable shaking time $\ts$, the BEC is released and the center of mass of the atomic density distribution is recorded after a time of flight $t_\text{ToF}=140$~ms, which reflects the {\it in-situ} center-of-mass velocity $v(\ts)$ along ${\bf e}_z$.

In Fig.~\ref{Fig2}(c), we show examples of $v(\ts)$ for three drive frequencies (below, on, and above resonance), and for two drive amplitudes ($\Us/\kB=0.2$ and $1.6$ nK). While at early times transient dynamics are visible, a monochromatic steady state is established at later times. We fit the data for $\ts > 0.6$~s with $v(\ts)=v_{\textrm{m}} \sin(\omega \ts+\phi)$ (solid lines) and extract the amplitude $v_{\textrm{m}}$ and phase $\phi$ of the velocity response. For the weaker drive ($\Us/\mu\approx 0.1$) we observe significant response only on resonance. However, for the stronger drive ($\Us/\mu\approx 0.7$) we observe comparable response amplitudes at all three frequencies, indicating significant broadening, a clear signature of nonlinear behavior.

To characterize this nonlinear behavior we plot $v_\textrm{m}$ as a function of $\Us$ on log-log scale for various drive frequencies in Fig.~\ref{Fig2}(d).
For linear response, with amplitude-independent damping, one would get $v_\textrm{m}\propto \Us$, for all $\omega$. 
Instead, on resonance ($\omega=\omega_d$) we observe power-law behavior $v_\textrm{m}\propto \Us^{0.32(1)}$, reminiscent of classical models with a cubic nonlinear damping [which generically predict $v_\textrm{m}\propto \Us^{1/3}$ on resonance~(see Section~I in \cite{Supplementary})].
Away from resonance, $v_\textrm{m}$ is smaller than on resonance (for the same $\Us$), but for stronger drives a progressively broader range of frequencies becomes effectively resonant.

\begin{figure}
	\centerline{\includegraphics[width=\columnwidth]{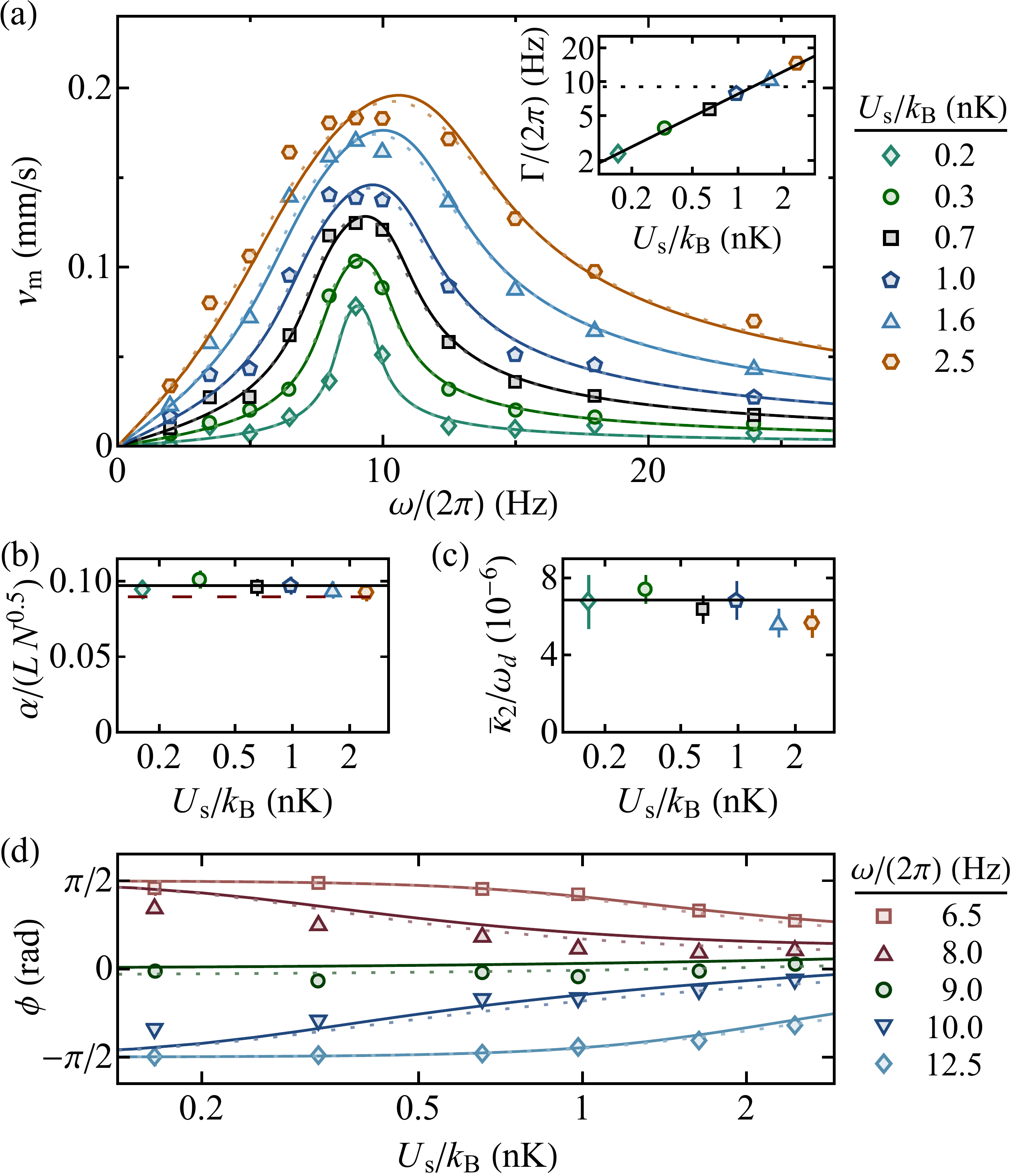}}
	\caption{Nonlinear response functions and comparison with theory. (a) Steady-state velocity response amplitude $v_\textrm{m}$ as a function of $\omega$, for six different drive strengths $U_\textrm{s}$ (see legend), alongside the corresponding fits to our model (see text for details).
		The inset shows $\Gamma$, the extracted full width at half maximum of $v_\textrm{m}(\omega)$, as a function of $\Us$ on log-log scale, while the horizontal dotted line indicates $\omega_d$.
	The solid line is a fit to the data, $\Gamma\propto \Us^{0.67(2)}$.
	For reference, our experimental upper bound on linear damping rate is $2\pi \times 0.3$~s$^{-1}$  \cite{garratt2019from}.
	(b,c) Extracted model fit parameters for the response curves $v_\textrm{m}(\omega)$ in (a). The dashed line in (b) shows a theoretical estimate (see text), while the solid lines in (b,c) depict the average for $\Us/k_\textrm{B}<1$~nK.
	(d) Phase response as a function of $\Us$. The solid lines show the prediction of the model with the extracted average parameters [solid lines in (b,c)] for $\text{Re}[\kappa_2]=0$. The dotted lines instead show corresponding predictions which include a small nonzero $\text{Re}[\kappa_2]$ (see text). Where not visible, the fitting error bars are smaller than the point size.
	} \label{Fig3}
\end{figure}

In the following we introduce a universal theoretical model for this nonlinear behavior, and compare it to the full experimental response curves $v_{\rm m}(\omega)$ and $\phi(\omega)$ (see Fig.~\ref{Fig3}).
Within quantum theory, in the spirit of Fig.~\ref{Fig1}(b), we associate the cube-root scaling $v_\textrm{m}\propto \Us^{1/3}$ with the decay of the mode towards higher energies via a two-body process~\footnote{More generally, a $M$-body process of this form would lead to a $1/(2M-1)$ power-law scaling (see Section~VI in \cite{Supplementary}).}, resulting in a decay rate proportional to the number of phonons. Once excited, this `secondary' mode couples to other modes, which act as a quasi-continuum of states to which the excitation can decay.
We formalize this picture by developing a simple model in which the fundamental excitation of energy $\hbar\omega_d$ is created by an external forcing, and is coupled to an auxiliary mode of energy $\hbar\omega_b$~(see Sections II-V in \cite{Supplementary}).
The Hamiltonian describing the system is
\begin{eqnarray}
\label{eq:ham}
\hat{H} &=& \hbar\omega _{d}\hat{d}^{\dag }\hat{d}+ \hbar\omega _{b}\hat{b}^{\dag }\hat{b}%
+\hbar\lambda ( \hat{b}^{\dag }\hat{d}\hat{d}+\text{H.c.})  \nonumber\\
&+& \hbar \Omega\sin \left( \omega t \right) ( \hat{d}^{\dag}+\hat{d})\,,
\end{eqnarray}
where $\hat{d}^\dag$ ($\hat{d}$) and $\hat{b}^\dag$ ($\hat{b}$) are, respectively, the creation (annihilation) operators for the fundamental and the auxiliary mode, $\lambda$ is the coupling strength, and $\Omega$ is the strength of the drive. 
We incorporate the decay of the auxiliary mode into the quasi-continuum via a master equation approach.
By tracing out the auxiliary mode, we derive, within a mean-field approximation, an equation of motion for the mean dipole $d(t)\equiv\langle{\hat{d}}(t)\rangle$:
\begin{equation}
\label{eq:dipole}
i\partial _{t}d-(\omega_{d}+\kappa_{2}|d|^{2})d=\Omega \sin(\omega t)\,.
\end{equation}
The generally complex $\kappa_{2}$ captures the nonlinear effects to leading order. Specifically, $\text{Re}[\kappa_2]$ and $\text{Im}[\kappa_2]$ correspond, respectively, to a frequency shift (due to the self-interaction) and a nonlinear damping (due to the mediated coupling to the continuum); expressions for $\kappa_2$ in terms of the microscopic model parameters are provided in~Section~III in \cite{Supplementary}.
In practice, $\kappa_2$ is sensitive to the details of the trapping potential, and it is more convenient to extract it directly from the experimental data.
However, the form of Eq.~(\ref{eq:dipole}) is universal in that it does not depend on the exact loss mechanism of the auxiliary mode, nor the number of auxiliary excitations involved in the elementary interaction process [see Eq.~(\ref{eq:ham})].

%%%%%%%%%%%%%%%%%%%%%%%%%%%%%%%%%%%%%%%%%%
%%%%%%%%%%%%%%%%%%%%%%%%%%%%%%%%%%%%%%%%%%% 

To compare our experimental data to the theory, we relate $d(t)$ to the main experimental observable $v(t)=(2\alpha/N)\partial_t \textrm{Re}[d(t)]$, where $\alpha$ is the matrix element of the position operator $\hat{z}$ between the ground state and the lowest-lying excitation (see Section~IV in \cite{Supplementary}).
In terms of the experimental parameters, the drive amplitude in Eq.~(\ref{eq:ham}) is $\Omega=\alpha \Us/(\hbar L)$. 
We determine the parameters of the model by fitting the $v_\textrm{m}(\omega)$ response curves to the steady-state numerical solutions of Eq.~(\ref{eq:dipole}) for each $\Us$. The resulting fits are shown in Fig.~\ref{Fig3}(a) as solid lines, where for simplicity we first assume that the nonlinear coefficient $\kappa_2$ is imaginary so that the adjustable parameters are $\bar{\kappa}_2\equiv i\kappa_2$ and $\alpha$.
We see that for $\Us\lesssim \kB\times 2$~nK the fitted model captures the experimental data well.

Only for $\Us \gtrsim \kB\times2$~nK do deviations between the model and the data become apparent. In the inset of Fig.~\ref{Fig3}(a) we plot the extracted full width at half maximum of the spectral lines, $\Gamma$, as a function of $\Us$, which reveals that the deviation between the model and the data occurs only once $\Gamma\gtrsim\omega_d$. For $\Us/\kB = 1.6$~nK, we estimate that $\bar{\kappa}_2 |d|^2 \approx 0.5\omega_d$, and higher-order nonlinearities could become important.

The parameters extracted from each $v_\textrm{m}(\omega)$ curve are shown in Figs.~\ref{Fig3}(b,c). Crucially, both $\alpha$ and $\bar{\kappa}_2$ do not depend on $\Us$ within experimental errors, demonstrating that the model [Eq.~\eqref{eq:dipole}] captures the nonlinear $\Us$-dependent response. Averaging the fitted parameters well within the range of validity of the model ($\Us/\kB<1$~nK) we obtain $\alpha=0.097(3) L \sqrt{N}$ and $\bar{\kappa}_2/\omega_d= 6.9(4) \times 10^{-6}$ per phonon (solid horizontal lines).
A calculation assuming a cylindrical-box-trapped BEC in the Thomas--Fermi regime~(Section~III in \cite{Supplementary}) yields $\alpha/(L\sqrt{N})=2^{5/4}/\pi^{3/2}\sqrt{\xi/L}\approx 0.090$~[dashed line in Fig.~\ref{Fig3}(b)], in good agreement with the measurement.
 
 Furthermore, we find that $\Gamma\propto \Us^{0.67(2)}$, which interestingly suggests that even though $v_\textrm{m}/\Us$ and $\Gamma$ are not independent of $\Us$ as for a linearly-damped harmonic oscillator, one still recovers $v_\textrm{m}\Gamma \propto \Us$.
 Naively, this reflects the energy balance condition for a driven-dissipative steady state; $U_{\textrm{s}}v_\textrm{m}$ is the driving power, and $v^2_\textrm{m}\Gamma$ is akin to the energy dissipation rate, the energy stored in the system being $\propto v^2_\textrm{m}$.
 
 %%%%%%%%%%%%%%%%%%%%%%%%%%%%%%%%%%%%%%%%%%
\begin{figure}[t]
	\centerline{\includegraphics[width=\columnwidth]{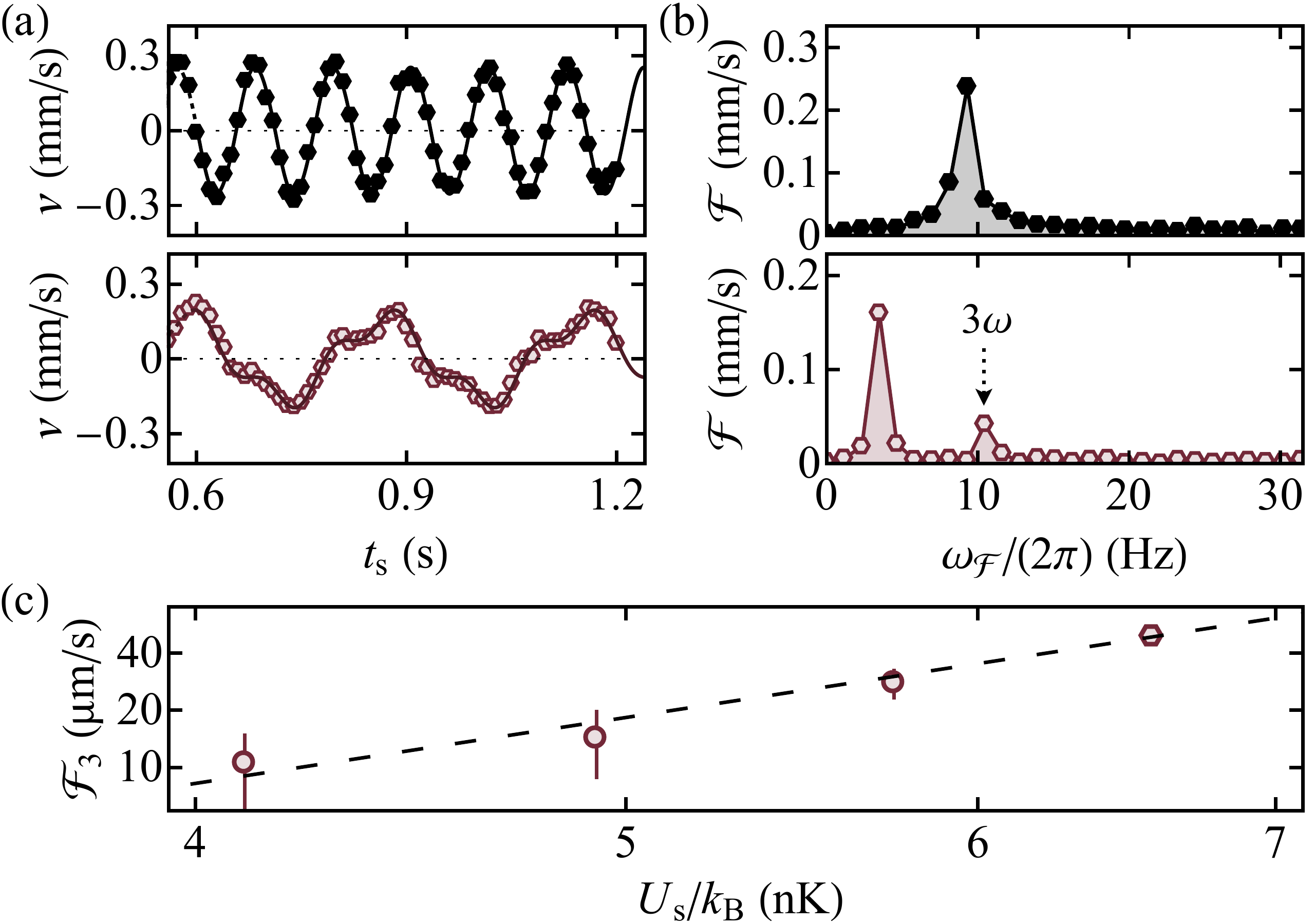}}
	\caption{Observation of third-harmonic generation. (a) Steady-state center-of-mass velocity response $v(\ts)$ for a strong drive ($\Us/\kB=6.6$~nK) at $\omega/(2\pi)= 9$~Hz (top) and $\omega/(2\pi)= 3.5$~Hz (bottom). In the latter case we clearly observe a non-monochromatic response. The solid lines show fits to the data~\cite{nldamp-FN-THG}. (b) Corresponding Fourier spectra $\mathcal{F}$, which for $\omega/(2\pi)= 3.5$~Hz reveals a response at $3\omega$. (c) Extracted third-harmonic generation amplitude $\mathcal{F}_3$  for $\omega/(2\pi)=3.5$~Hz plotted as a function of $\Us$ on log-log scale. The dashed line shows a power-law fit, which gives $\mathcal{F}_3\propto \Us^{3.6\pm1.2}$, consistent with the cubic scaling behavior characteristic of third-harmonic generation.
	} \label{Fig4}
\end{figure}
%%%%%%%%%%%%%%%%%%%%%%%%%%%%%%%%%%%%%%%%%%% 
 
In Fig.~\ref{Fig3}(d) we show the complementary data on the phase of the \emph{velocity} response, $\phi$, as a function of $\Us$ for various drive frequencies.
As expected, $\phi$ is $\pi/2$ (-$\pi/2$) out of phase with the drive below (above) resonance.
Interestingly, for low drive strengths ($\Us/\kB < 1~$nK) the on-resonance $\phi(\omega_d)$ is observably nonzero and nearly independent of $\Us$. Averaging the data for $\Us/\kB<1$~nK gives $\phi=-0.12(4)$, which suggests that $\text{Re}[\kappa_2]=0.12(4)\text{Im}[\kappa_2]$ within our model [see Section~IV in \cite{Supplementary}); this \emph{a posteriori} justifies our initial assumption that $\lvert\text{Re}[\kappa_2]/\text{Im}[\kappa_2]\lvert\ll 1$. Including this small nonzero $\text{Re}[\kappa_2]$ in the model further improves agreement with the data [dotted lines in Figs.~\ref{Fig3}(a,d)], but the difference is small.

For a nonlinear response one also generically expects the possibility of harmonic generation. 
Experimentally, we did not observe evidence of higher-harmonic generation for $\Us/\kB < 3$~nK, so in the final part of this Letter we turn to even larger drive amplitudes (up to $\Us/\kB=6.6$~nK).
For most driving frequencies $\omega$ we only observe a monochromatic response at the drive frequency, even for strong drives (see top panels in Figs.~\ref{Fig4}(a,b) where $\omega/(2\pi)=9$~Hz~\footnote{In this case the response amplitude still follows the resonant scaling $v_\textrm{m}\propto \Us^{1/3}$ seen in Fig.~\ref{Fig2}(b).}).
However, the behavior is markedly different for $\omega/(2\pi)= 3.5$~Hz (bottom panels), for which $3 \omega$ is close to the resonant $\omega_d$.
In this case the Fourier spectrum of the time-domain response [Fig.~\ref{Fig4}(b)] shows a clear peak at $3 \omega$, signalling third-harmonic generation, in addition to the main peak at $\omega$.
In Fig.~\ref{Fig4}(c) we plot the third-harmonic amplitude $\mathcal{F}_3$ versus $\Us$ on log-log scale~\cite{nldamp-FN-THG}.
A power-law fit (dashed line) gives $\mathcal{F}_3 \propto \Us^{3.6\pm1.2}$, consistent with a cubic scaling in $\Us$ (characteristic of third-harmonic generation) and in qualitative agreement with predictions from Eq.~(\ref{eq:dipole})~(Section~V in \cite{Supplementary}).
 
In conclusion, we have experimentally and theoretically studied the nonlinear decay of the fundamental gapped excitation of a Bose--Einstein condensate. Our experiments reveal a cubic damping mechanism as well as third-harmonic generation. To understand these nonlinear effects we develop a mean-field model based on a microscopic theory at lowest order.
Our work paves the way to a microscopic understanding of the genesis of a turbulent cascade~\cite{navon2016emergence,navon2019synthetic}, when the energy leakage from the driven lowest-lying mode is sufficiently large to sustain a non-equilibrium steady state. 
The decay of gapped excitations is an understudied but generic problem in quantum many-body physics, and our approach could be extended to other systems, where it could offer insight into the interactions between those elementary excitations.

We thank Robert P. Smith, Maciej Ga\l{}ka, and Nishant Dogra for helpful discussions and comments on the manuscript. This work was supported by EPSRC [Grant Nos.~EP/N011759/1, EP/P009565/1, EP/N011759/1, EP/P009565/1 and EP/K030094/1], ERC (QBox), QuantERA (NAQUAS, EPSRC Grant No.~EP/R043396/1), AFOSR, and ARO. N.~N. acknowledges support from Trinity College (Cambridge), the David and Lucile Packard Foundation, the Alfred P. Sloan Foundation, and NSF CAREER (1945324). C.~E. acknowledges support from Jesus College (Cambridge). T.~A.~H. acknowledges support from the EU Marie Sk\l{}odowska-Curie program [Grant No.~MSCA-IF- 2018 840081]. Z.~H. acknowledges support from the Royal Society Wolfson Fellowship. N.~R.~C. acknowledges the support of the Simons Foundation.

\bibliographystyle{apsrev4-1}

\begin{thebibliography}{39}%
\makeatletter
\providecommand \@ifxundefined [1]{%
 \@ifx{#1\undefined}
}%
\providecommand \@ifnum [1]{%
 \ifnum #1\expandafter \@firstoftwo
 \else \expandafter \@secondoftwo
 \fi
}%
\providecommand \@ifx [1]{%
 \ifx #1\expandafter \@firstoftwo
 \else \expandafter \@secondoftwo
 \fi
}%
\providecommand \natexlab [1]{#1}%
\providecommand \enquote  [1]{``#1''}%
\providecommand \bibnamefont  [1]{#1}%
\providecommand \bibfnamefont [1]{#1}%
\providecommand \citenamefont [1]{#1}%
\providecommand \href@noop [0]{\@secondoftwo}%
\providecommand \href [0]{\begingroup \@sanitize@url \@href}%
\providecommand \@href[1]{\@@startlink{#1}\@@href}%
\providecommand \@@href[1]{\endgroup#1\@@endlink}%
\providecommand \@sanitize@url [0]{\catcode `\\12\catcode `\$12\catcode
  `\&12\catcode `\#12\catcode `\^12\catcode `\_12\catcode `\%12\relax}%
\providecommand \@@startlink[1]{}%
\providecommand \@@endlink[0]{}%
\providecommand \url  [0]{\begingroup\@sanitize@url \@url }%
\providecommand \@url [1]{\endgroup\@href {#1}{\urlprefix }}%
\providecommand \urlprefix  [0]{URL }%
\providecommand \Eprint [0]{\href }%
\providecommand \doibase [0]{https://doi.org/}%
\providecommand \selectlanguage [0]{\@gobble}%
\providecommand \bibinfo  [0]{\@secondoftwo}%
\providecommand \bibfield  [0]{\@secondoftwo}%
\providecommand \translation [1]{[#1]}%
\providecommand \BibitemOpen [0]{}%
\providecommand \bibitemStop [0]{}%
\providecommand \bibitemNoStop [0]{.\EOS\space}%
\providecommand \EOS [0]{\spacefactor3000\relax}%
\providecommand \BibitemShut  [1]{\csname bibitem#1\endcsname}%
\let\auto@bib@innerbib\@empty
%</preamble>

\item[$^{\color{myColor}\ast}$] These two authors contributed equally.
\item[$^{\color{myColor}\dag}$] To whom correspondence should be addressed:\\ ce330@cam.ac.uk, zhengwei8796@foxmail.com.
\bibitem [{\citenamefont {Fetter}\ and\ \citenamefont
  {Walecka}(2003)}]{fetter2012quantum}%
  \BibitemOpen
  \bibfield  {author} {\bibinfo {author} {\bibfnamefont {A.~L.}\ \bibnamefont
  {Fetter}}\ and\ \bibinfo {author} {\bibfnamefont {J.~D.}\ \bibnamefont
  {Walecka}},\ }\href@noop {} {\emph {\bibinfo {title} {Quantum Theory of
  Many-Particle Systems}}}\ (\bibinfo  {publisher} {Dover Publications},\
  \bibinfo {year} {2003})\BibitemShut {NoStop}%
\bibitem [{\citenamefont {Henley}\ and\ \citenamefont
  {Wilets}(1964)}]{henley1964energy}%
  \BibitemOpen
  \bibfield  {author} {\bibinfo {author} {\bibfnamefont {E.~M.}\ \bibnamefont
  {Henley}}\ and\ \bibinfo {author} {\bibfnamefont {L.}~\bibnamefont
  {Wilets}},\ }\bibfield  {title} {\bibinfo {title} {{Energy Gap in Nuclear
  Matter. I. Extended Theory}},\ }\href
  {https://doi.org/10.1103/PhysRev.133.B1118} {\bibfield  {journal} {\bibinfo
  {journal} {Phys. Rev.}\ }\textbf {\bibinfo {volume} {133}},\ \bibinfo {pages}
  {B1118} (\bibinfo {year} {1964})}\BibitemShut {NoStop}%
\bibitem [{\citenamefont {Ashoori}(1996)}]{ashoori1996electrons}%
  \BibitemOpen
  \bibfield  {author} {\bibinfo {author} {\bibfnamefont {R.~C.}\ \bibnamefont
  {Ashoori}},\ }\bibfield  {title} {\bibinfo {title} {{Electrons in artificial
  atoms}},\ }\href {https://doi.org/10.1038/379413a0} {\bibfield  {journal}
  {\bibinfo  {journal} {Nature}\ }\textbf {\bibinfo {volume} {379}},\ \bibinfo
  {pages} {413} (\bibinfo {year} {1996})}\BibitemShut {NoStop}%
\bibitem [{\citenamefont {Cooper}(1956)}]{cooper1956bound}%
  \BibitemOpen
  \bibfield  {author} {\bibinfo {author} {\bibfnamefont {L.~N.}\ \bibnamefont
  {Cooper}},\ }\bibfield  {title} {\bibinfo {title} {{Bound Electron Pairs in a
  Degenerate Fermi Gas}},\ }\href {https://doi.org/10.1103/PhysRev.104.1189}
  {\bibfield  {journal} {\bibinfo  {journal} {Phys. Rev.}\ }\textbf {\bibinfo
  {volume} {104}},\ \bibinfo {pages} {1189} (\bibinfo {year}
  {1956})}\BibitemShut {NoStop}%
\bibitem [{\citenamefont {Kohn}(1964)}]{kohn1964theory}%
  \BibitemOpen
  \bibfield  {author} {\bibinfo {author} {\bibfnamefont {W.}~\bibnamefont
  {Kohn}},\ }\bibfield  {title} {\bibinfo {title} {{Theory of the Insulating
  State}},\ }\href {https://doi.org/10.1103/PhysRev.133.A171} {\bibfield
  {journal} {\bibinfo  {journal} {Phys. Rev.}\ }\textbf {\bibinfo {volume}
  {133}},\ \bibinfo {pages} {A171} (\bibinfo {year} {1964})}\BibitemShut
  {NoStop}%
\bibitem [{\citenamefont {Garc\'{\i}a-Garc\'{\i}a}\ \emph
  {et~al.}(2008)\citenamefont {Garc\'{\i}a-Garc\'{\i}a}, \citenamefont
  {Urbina}, \citenamefont {Yuzbashyan}, \citenamefont {Richter},\ and\
  \citenamefont {Altshuler}}]{garcia2008bardeen}%
  \BibitemOpen
  \bibfield  {author} {\bibinfo {author} {\bibfnamefont {A.~M.}\ \bibnamefont
  {Garc\'{\i}a-Garc\'{\i}a}}, \bibinfo {author} {\bibfnamefont {J.~D.}\
  \bibnamefont {Urbina}}, \bibinfo {author} {\bibfnamefont {E.~A.}\
  \bibnamefont {Yuzbashyan}}, \bibinfo {author} {\bibfnamefont
  {K.}~\bibnamefont {Richter}},\ and\ \bibinfo {author} {\bibfnamefont {B.~L.}\
  \bibnamefont {Altshuler}},\ }\bibfield  {title} {\bibinfo {title}
  {{Bardeen--Cooper--Schrieffer Theory of Finite-Size Superconducting Metallic
  Grains}},\ }\href {https://doi.org/10.1103/PhysRevLett.100.187001} {\bibfield
   {journal} {\bibinfo  {journal} {Phys. Rev. Lett.}\ }\textbf {\bibinfo
  {volume} {100}},\ \bibinfo {pages} {187001} (\bibinfo {year}
  {2008})}\BibitemShut {NoStop}%
\bibitem [{\citenamefont {Jin}\ \emph {et~al.}(1996)\citenamefont {Jin},
  \citenamefont {Ensher}, \citenamefont {Matthews}, \citenamefont {Wieman},\
  and\ \citenamefont {Cornell}}]{jin1996collective}%
  \BibitemOpen
  \bibfield  {author} {\bibinfo {author} {\bibfnamefont {D.~S.}\ \bibnamefont
  {Jin}}, \bibinfo {author} {\bibfnamefont {J.~R.}\ \bibnamefont {Ensher}},
  \bibinfo {author} {\bibfnamefont {M.~R.}\ \bibnamefont {Matthews}}, \bibinfo
  {author} {\bibfnamefont {C.~E.}\ \bibnamefont {Wieman}},\ and\ \bibinfo
  {author} {\bibfnamefont {E.~A.}\ \bibnamefont {Cornell}},\ }\bibfield
  {title} {\bibinfo {title} {{Collective Excitations of a Bose--Einstein
  Condensate in a Dilute Gas}},\ }\href
  {https://doi.org/10.1103/PhysRevLett.77.420} {\bibfield  {journal} {\bibinfo
  {journal} {Phys. Rev. Lett.}\ }\textbf {\bibinfo {volume} {77}},\ \bibinfo
  {pages} {420} (\bibinfo {year} {1996})}\BibitemShut {NoStop}%
\bibitem [{\citenamefont {Mewes}\ \emph {et~al.}(1996)\citenamefont {Mewes},
  \citenamefont {Andrews}, \citenamefont {van Druten}, \citenamefont {Kurn},
  \citenamefont {Durfee}, \citenamefont {Townsend},\ and\ \citenamefont
  {Ketterle}}]{mewes1996collective}%
  \BibitemOpen
  \bibfield  {author} {\bibinfo {author} {\bibfnamefont {M.-O.}\ \bibnamefont
  {Mewes}}, \bibinfo {author} {\bibfnamefont {M.~R.}\ \bibnamefont {Andrews}},
  \bibinfo {author} {\bibfnamefont {N.~J.}\ \bibnamefont {van Druten}},
  \bibinfo {author} {\bibfnamefont {D.~M.}\ \bibnamefont {Kurn}}, \bibinfo
  {author} {\bibfnamefont {D.~S.}\ \bibnamefont {Durfee}}, \bibinfo {author}
  {\bibfnamefont {C.~G.}\ \bibnamefont {Townsend}},\ and\ \bibinfo {author}
  {\bibfnamefont {W.}~\bibnamefont {Ketterle}},\ }\bibfield  {title} {\bibinfo
  {title} {{Collective Excitations of a {B}ose--{E}instein Condensate in a
  Magnetic Trap}},\ }\href {https://doi.org/10.1103/PhysRevLett.77.988}
  {\bibfield  {journal} {\bibinfo  {journal} {Phys. Rev. Lett.}\ }\textbf
  {\bibinfo {volume} {77}},\ \bibinfo {pages} {988} (\bibinfo {year}
  {1996})}\BibitemShut {NoStop}%
\bibitem [{\citenamefont {Andrews}\ \emph {et~al.}(1997)\citenamefont
  {Andrews}, \citenamefont {Kurn}, \citenamefont {Miesner}, \citenamefont
  {Durfee}, \citenamefont {Townsend}, \citenamefont {Inouye},\ and\
  \citenamefont {Ketterle}}]{andrews1997propagation}%
  \BibitemOpen
  \bibfield  {author} {\bibinfo {author} {\bibfnamefont {M.~R.}\ \bibnamefont
  {Andrews}}, \bibinfo {author} {\bibfnamefont {D.~M.}\ \bibnamefont {Kurn}},
  \bibinfo {author} {\bibfnamefont {H.-J.}\ \bibnamefont {Miesner}}, \bibinfo
  {author} {\bibfnamefont {D.~S.}\ \bibnamefont {Durfee}}, \bibinfo {author}
  {\bibfnamefont {C.~G.}\ \bibnamefont {Townsend}}, \bibinfo {author}
  {\bibfnamefont {S.}~\bibnamefont {Inouye}},\ and\ \bibinfo {author}
  {\bibfnamefont {W.}~\bibnamefont {Ketterle}},\ }\bibfield  {title} {\bibinfo
  {title} {{Propagation of Sound in a {B}ose--{E}instein Condensate}},\ }\href
  {https://doi.org/10.1103/PhysRevLett.79.553} {\bibfield  {journal} {\bibinfo
  {journal} {Phys. Rev. Lett.}\ }\textbf {\bibinfo {volume} {79}},\ \bibinfo
  {pages} {553} (\bibinfo {year} {1997})}\BibitemShut {NoStop}%
\bibitem [{\citenamefont {Stamper-Kurn}\ \emph {et~al.}(1998)\citenamefont
  {Stamper-Kurn}, \citenamefont {Miesner}, \citenamefont {Inouye},
  \citenamefont {Andrews},\ and\ \citenamefont
  {Ketterle}}]{stamper1998collisionless}%
  \BibitemOpen
  \bibfield  {author} {\bibinfo {author} {\bibfnamefont {D.~M.}\ \bibnamefont
  {Stamper-Kurn}}, \bibinfo {author} {\bibfnamefont {H.-J.}\ \bibnamefont
  {Miesner}}, \bibinfo {author} {\bibfnamefont {S.}~\bibnamefont {Inouye}},
  \bibinfo {author} {\bibfnamefont {M.~R.}\ \bibnamefont {Andrews}},\ and\
  \bibinfo {author} {\bibfnamefont {W.}~\bibnamefont {Ketterle}},\ }\bibfield
  {title} {\bibinfo {title} {{Collisionless and Hydrodynamic Excitations of a
  {B}ose--{E}instein Condensate}},\ }\href
  {https://doi.org/https://doi.org/10.1103/PhysRevLett.81.500} {\bibfield
  {journal} {\bibinfo  {journal} {Phys. Rev. Lett.}\ }\textbf {\bibinfo
  {volume} {81}},\ \bibinfo {pages} {500} (\bibinfo {year} {1998})}\BibitemShut
  {NoStop}%
\bibitem [{\citenamefont {Stamper-Kurn}\ \emph {et~al.}(1999)\citenamefont
  {Stamper-Kurn}, \citenamefont {Chikkatur}, \citenamefont {G\"orlitz},
  \citenamefont {Inouye}, \citenamefont {Gupta}, \citenamefont {Pritchard},\
  and\ \citenamefont {Ketterle}}]{stamper1999excitation}%
  \BibitemOpen
  \bibfield  {author} {\bibinfo {author} {\bibfnamefont {D.~M.}\ \bibnamefont
  {Stamper-Kurn}}, \bibinfo {author} {\bibfnamefont {A.~P.}\ \bibnamefont
  {Chikkatur}}, \bibinfo {author} {\bibfnamefont {A.}~\bibnamefont
  {G\"orlitz}}, \bibinfo {author} {\bibfnamefont {S.}~\bibnamefont {Inouye}},
  \bibinfo {author} {\bibfnamefont {S.}~\bibnamefont {Gupta}}, \bibinfo
  {author} {\bibfnamefont {D.~E.}\ \bibnamefont {Pritchard}},\ and\ \bibinfo
  {author} {\bibfnamefont {W.}~\bibnamefont {Ketterle}},\ }\bibfield  {title}
  {\bibinfo {title} {{Excitation of Phonons in a {B}ose--{E}instein Condensate
  by Light Scattering}},\ }\href {https://doi.org/10.1103/PhysRevLett.83.2876}
  {\bibfield  {journal} {\bibinfo  {journal} {Phys. Rev. Lett.}\ }\textbf
  {\bibinfo {volume} {83}},\ \bibinfo {pages} {2876} (\bibinfo {year}
  {1999})}\BibitemShut {NoStop}%
\bibitem [{\citenamefont {Stenger}\ \emph {et~al.}(1999)\citenamefont
  {Stenger}, \citenamefont {Inouye}, \citenamefont {Chikkatur}, \citenamefont
  {Stamper-Kurn}, \citenamefont {Pritchard},\ and\ \citenamefont
  {Ketterle}}]{stenger1999bragg}%
  \BibitemOpen
  \bibfield  {author} {\bibinfo {author} {\bibfnamefont {J.}~\bibnamefont
  {Stenger}}, \bibinfo {author} {\bibfnamefont {S.}~\bibnamefont {Inouye}},
  \bibinfo {author} {\bibfnamefont {A.~P.}\ \bibnamefont {Chikkatur}}, \bibinfo
  {author} {\bibfnamefont {D.~M.}\ \bibnamefont {Stamper-Kurn}}, \bibinfo
  {author} {\bibfnamefont {D.~E.}\ \bibnamefont {Pritchard}},\ and\ \bibinfo
  {author} {\bibfnamefont {W.}~\bibnamefont {Ketterle}},\ }\bibfield  {title}
  {\bibinfo {title} {{Bragg Spectroscopy of a {B}ose--{E}instein Condensate}},\
  }\href {https://doi.org/10.1103/PhysRevLett.82.4569} {\bibfield  {journal}
  {\bibinfo  {journal} {Phys. Rev. Lett.}\ }\textbf {\bibinfo {volume} {82}},\
  \bibinfo {pages} {4569} (\bibinfo {year} {1999})}\BibitemShut {NoStop}%
\bibitem [{\citenamefont {Marag\`o}\ \emph {et~al.}(2000)\citenamefont
  {Marag\`o}, \citenamefont {Hopkins}, \citenamefont {Arlt}, \citenamefont
  {Hodby}, \citenamefont {Hechenblaikner},\ and\ \citenamefont
  {Foot}}]{marago2000observation}%
  \BibitemOpen
  \bibfield  {author} {\bibinfo {author} {\bibfnamefont {O.~M.}\ \bibnamefont
  {Marag\`o}}, \bibinfo {author} {\bibfnamefont {S.~A.}\ \bibnamefont
  {Hopkins}}, \bibinfo {author} {\bibfnamefont {J.}~\bibnamefont {Arlt}},
  \bibinfo {author} {\bibfnamefont {E.}~\bibnamefont {Hodby}}, \bibinfo
  {author} {\bibfnamefont {G.}~\bibnamefont {Hechenblaikner}},\ and\ \bibinfo
  {author} {\bibfnamefont {C.~J.}\ \bibnamefont {Foot}},\ }\bibfield  {title}
  {\bibinfo {title} {{Observation of the Scissors Mode and Evidence for
  Superfluidity of a Trapped Bose--Einstein Condensed Gas}},\ }\href
  {https://doi.org/10.1103/PhysRevLett.84.2056} {\bibfield  {journal} {\bibinfo
   {journal} {Phys. Rev. Lett.}\ }\textbf {\bibinfo {volume} {84}},\ \bibinfo
  {pages} {2056} (\bibinfo {year} {2000})}\BibitemShut {NoStop}%
\bibitem [{\citenamefont {Abo-Shaeer}\ \emph {et~al.}(2001)\citenamefont
  {Abo-Shaeer}, \citenamefont {Raman}, \citenamefont {Vogels},\ and\
  \citenamefont {Ketterle}}]{aboshaeer2001observation}%
  \BibitemOpen
  \bibfield  {author} {\bibinfo {author} {\bibfnamefont {J.~R.}\ \bibnamefont
  {Abo-Shaeer}}, \bibinfo {author} {\bibfnamefont {C.}~\bibnamefont {Raman}},
  \bibinfo {author} {\bibfnamefont {J.~M.}\ \bibnamefont {Vogels}},\ and\
  \bibinfo {author} {\bibfnamefont {W.}~\bibnamefont {Ketterle}},\ }\bibfield
  {title} {\bibinfo {title} {{Observation of Vortex Lattices in Bose--Einstein
  Condensates}},\ }\href {https://doi.org/10.1126/science.1060182} {\bibfield
  {journal} {\bibinfo  {journal} {Science}\ }\textbf {\bibinfo {volume}
  {292}},\ \bibinfo {pages} {476} (\bibinfo {year} {2001})}\BibitemShut
  {NoStop}%
\bibitem [{\citenamefont {Steinhauer}\ \emph {et~al.}(2002)\citenamefont
  {Steinhauer}, \citenamefont {Ozeri}, \citenamefont {Katz},\ and\
  \citenamefont {Davidson}}]{steinhauer2002excitation}%
  \BibitemOpen
  \bibfield  {author} {\bibinfo {author} {\bibfnamefont {J.}~\bibnamefont
  {Steinhauer}}, \bibinfo {author} {\bibfnamefont {R.}~\bibnamefont {Ozeri}},
  \bibinfo {author} {\bibfnamefont {N.}~\bibnamefont {Katz}},\ and\ \bibinfo
  {author} {\bibfnamefont {N.}~\bibnamefont {Davidson}},\ }\bibfield  {title}
  {\bibinfo {title} {{Excitation Spectrum of a {B}ose--{E}instein
  Condensate}},\ }\href {https://doi.org/10.1103/PhysRevLett.88.120407}
  {\bibfield  {journal} {\bibinfo  {journal} {Phys. Rev. Lett.}\ }\textbf
  {\bibinfo {volume} {88}},\ \bibinfo {pages} {120407} (\bibinfo {year}
  {2002})}\BibitemShut {NoStop}%
\bibitem [{\citenamefont {Vogels}\ \emph {et~al.}(2002)\citenamefont {Vogels},
  \citenamefont {Xu}, \citenamefont {Raman}, \citenamefont {Abo-Shaeer},\ and\
  \citenamefont {Ketterle}}]{vogels2002experimental}%
  \BibitemOpen
  \bibfield  {author} {\bibinfo {author} {\bibfnamefont {J.~M.}\ \bibnamefont
  {Vogels}}, \bibinfo {author} {\bibfnamefont {K.}~\bibnamefont {Xu}}, \bibinfo
  {author} {\bibfnamefont {C.}~\bibnamefont {Raman}}, \bibinfo {author}
  {\bibfnamefont {J.~R.}\ \bibnamefont {Abo-Shaeer}},\ and\ \bibinfo {author}
  {\bibfnamefont {W.}~\bibnamefont {Ketterle}},\ }\bibfield  {title} {\bibinfo
  {title} {{Experimental Observation of the {B}ogoliubov Transformation for a
  {B}ose--{E}instein Condensed Gas}},\ }\href
  {https://doi.org/10.1103/PhysRevLett.88.060402} {\bibfield  {journal}
  {\bibinfo  {journal} {Phys. Rev. Lett.}\ }\textbf {\bibinfo {volume} {88}},\
  \bibinfo {pages} {060402} (\bibinfo {year} {2002})}\BibitemShut {NoStop}%
\bibitem [{\citenamefont {Katz}\ \emph {et~al.}(2002)\citenamefont {Katz},
  \citenamefont {Steinhauer}, \citenamefont {Ozeri},\ and\ \citenamefont
  {Davidson}}]{katz2002beliaev}%
  \BibitemOpen
  \bibfield  {author} {\bibinfo {author} {\bibfnamefont {N.}~\bibnamefont
  {Katz}}, \bibinfo {author} {\bibfnamefont {J.}~\bibnamefont {Steinhauer}},
  \bibinfo {author} {\bibfnamefont {R.}~\bibnamefont {Ozeri}},\ and\ \bibinfo
  {author} {\bibfnamefont {N.}~\bibnamefont {Davidson}},\ }\bibfield  {title}
  {\bibinfo {title} {{Beliaev Damping of Quasiparticles in a Bose--Einstein
  Condensate}},\ }\href {https://doi.org/10.1103/PhysRevLett.89.220401}
  {\bibfield  {journal} {\bibinfo  {journal} {Phys. Rev. Lett.}\ }\textbf
  {\bibinfo {volume} {89}},\ \bibinfo {pages} {220401} (\bibinfo {year}
  {2002})}\BibitemShut {NoStop}%
\bibitem [{\citenamefont {Khaykovich}\ \emph {et~al.}(2002)\citenamefont
  {Khaykovich}, \citenamefont {Schreck}, \citenamefont {Ferrari}, \citenamefont
  {Bourdel}, \citenamefont {Cubizolles}, \citenamefont {Carr}, \citenamefont
  {Castin},\ and\ \citenamefont {Salomon}}]{khaykovich2002formation}%
  \BibitemOpen
  \bibfield  {author} {\bibinfo {author} {\bibfnamefont {L.}~\bibnamefont
  {Khaykovich}}, \bibinfo {author} {\bibfnamefont {F.}~\bibnamefont {Schreck}},
  \bibinfo {author} {\bibfnamefont {G.}~\bibnamefont {Ferrari}}, \bibinfo
  {author} {\bibfnamefont {T.}~\bibnamefont {Bourdel}}, \bibinfo {author}
  {\bibfnamefont {J.}~\bibnamefont {Cubizolles}}, \bibinfo {author}
  {\bibfnamefont {L.~D.}\ \bibnamefont {Carr}}, \bibinfo {author}
  {\bibfnamefont {Y.}~\bibnamefont {Castin}},\ and\ \bibinfo {author}
  {\bibfnamefont {C.}~\bibnamefont {Salomon}},\ }\bibfield  {title} {\bibinfo
  {title} {{Formation of a Matter-Wave Bright Soliton}},\ }\href
  {https://doi.org/10.1126/science.1071021} {\bibfield  {journal} {\bibinfo
  {journal} {Science}\ }\textbf {\bibinfo {volume} {296}},\ \bibinfo {pages}
  {1290} (\bibinfo {year} {2002})}\BibitemShut {NoStop}%
\bibitem [{\citenamefont {Strecker}\ \emph {et~al.}(2002)\citenamefont
  {Strecker}, \citenamefont {Partridge}, \citenamefont {Truscott},\ and\
  \citenamefont {Hulet}}]{strecker2002formation}%
  \BibitemOpen
  \bibfield  {author} {\bibinfo {author} {\bibfnamefont {K.~E.}\ \bibnamefont
  {Strecker}}, \bibinfo {author} {\bibfnamefont {G.~B.}\ \bibnamefont
  {Partridge}}, \bibinfo {author} {\bibfnamefont {A.~G.}\ \bibnamefont
  {Truscott}},\ and\ \bibinfo {author} {\bibfnamefont {R.~G.}\ \bibnamefont
  {Hulet}},\ }\bibfield  {title} {\bibinfo {title} {{Formation and propagation
  of matter-wave soliton trains}},\ }\href {https://doi.org/10.1038/nature747}
  {\bibfield  {journal} {\bibinfo  {journal} {Nature}\ }\textbf {\bibinfo
  {volume} {417}},\ \bibinfo {pages} {150} (\bibinfo {year}
  {2002})}\BibitemShut {NoStop}%
\bibitem [{\citenamefont {Coddington}\ \emph {et~al.}(2003)\citenamefont
  {Coddington}, \citenamefont {Engels}, \citenamefont {Schweikhard},\ and\
  \citenamefont {Cornell}}]{coddington2003observation}%
  \BibitemOpen
  \bibfield  {author} {\bibinfo {author} {\bibfnamefont {I.}~\bibnamefont
  {Coddington}}, \bibinfo {author} {\bibfnamefont {P.}~\bibnamefont {Engels}},
  \bibinfo {author} {\bibfnamefont {V.}~\bibnamefont {Schweikhard}},\ and\
  \bibinfo {author} {\bibfnamefont {E.~A.}\ \bibnamefont {Cornell}},\
  }\bibfield  {title} {\bibinfo {title} {{Observation of Tkachenko Oscillations
  in Rapidly Rotating Bose--Einstein Condensates}},\ }\href
  {https://doi.org/10.1103/PhysRevLett.91.100402} {\bibfield  {journal}
  {\bibinfo  {journal} {Phys. Rev. Lett.}\ }\textbf {\bibinfo {volume} {91}},\
  \bibinfo {pages} {100402} (\bibinfo {year} {2003})}\BibitemShut {NoStop}%
\bibitem [{\citenamefont {Bretin}\ \emph {et~al.}(2003)\citenamefont {Bretin},
  \citenamefont {Rosenbusch}, \citenamefont {Chevy}, \citenamefont
  {Shlyapnikov},\ and\ \citenamefont {Dalibard}}]{bretin2003quadrupole}%
  \BibitemOpen
  \bibfield  {author} {\bibinfo {author} {\bibfnamefont {V.}~\bibnamefont
  {Bretin}}, \bibinfo {author} {\bibfnamefont {P.}~\bibnamefont {Rosenbusch}},
  \bibinfo {author} {\bibfnamefont {F.}~\bibnamefont {Chevy}}, \bibinfo
  {author} {\bibfnamefont {G.~V.}\ \bibnamefont {Shlyapnikov}},\ and\ \bibinfo
  {author} {\bibfnamefont {J.}~\bibnamefont {Dalibard}},\ }\bibfield  {title}
  {\bibinfo {title} {{Quadrupole Oscillation of a Single-Vortex Bose--Einstein
  Condensate: Evidence for Kelvin Modes}},\ }\href
  {https://doi.org/10.1103/PhysRevLett.90.100403} {\bibfield  {journal}
  {\bibinfo  {journal} {Phys. Rev. Lett.}\ }\textbf {\bibinfo {volume} {90}},\
  \bibinfo {pages} {100403} (\bibinfo {year} {2003})}\BibitemShut {NoStop}%
\bibitem [{\citenamefont {Engels}\ \emph {et~al.}(2007)\citenamefont {Engels},
  \citenamefont {Atherton},\ and\ \citenamefont
  {Hoefer}}]{engels2007observation}%
  \BibitemOpen
  \bibfield  {author} {\bibinfo {author} {\bibfnamefont {P.}~\bibnamefont
  {Engels}}, \bibinfo {author} {\bibfnamefont {C.}~\bibnamefont {Atherton}},\
  and\ \bibinfo {author} {\bibfnamefont {M.~A.}\ \bibnamefont {Hoefer}},\
  }\bibfield  {title} {\bibinfo {title} {{Observation of Faraday Waves in a
  Bose--Einstein Condensate}},\ }\href
  {https://doi.org/10.1103/PhysRevLett.98.095301} {\bibfield  {journal}
  {\bibinfo  {journal} {Phys. Rev. Lett.}\ }\textbf {\bibinfo {volume} {98}},\
  \bibinfo {pages} {095301} (\bibinfo {year} {2007})}\BibitemShut {NoStop}%
\bibitem [{\citenamefont {Papp}\ \emph {et~al.}(2008)\citenamefont {Papp},
  \citenamefont {Pino}, \citenamefont {Wild}, \citenamefont {Ronen},
  \citenamefont {Wieman}, \citenamefont {Jin},\ and\ \citenamefont
  {Cornell}}]{papp2008bragg}%
  \BibitemOpen
  \bibfield  {author} {\bibinfo {author} {\bibfnamefont {S.~B.}\ \bibnamefont
  {Papp}}, \bibinfo {author} {\bibfnamefont {J.~M.}\ \bibnamefont {Pino}},
  \bibinfo {author} {\bibfnamefont {R.~J.}\ \bibnamefont {Wild}}, \bibinfo
  {author} {\bibfnamefont {S.}~\bibnamefont {Ronen}}, \bibinfo {author}
  {\bibfnamefont {C.~E.}\ \bibnamefont {Wieman}}, \bibinfo {author}
  {\bibfnamefont {D.~S.}\ \bibnamefont {Jin}},\ and\ \bibinfo {author}
  {\bibfnamefont {E.~A.}\ \bibnamefont {Cornell}},\ }\bibfield  {title}
  {\bibinfo {title} {{Bragg Spectroscopy of a Strongly Interacting
  $^{85}\mathrm{Rb}$ Bose--Einstein Condensate}},\ }\href
  {https://doi.org/10.1103/PhysRevLett.101.135301} {\bibfield  {journal}
  {\bibinfo  {journal} {Phys. Rev. Lett.}\ }\textbf {\bibinfo {volume} {101}},\
  \bibinfo {pages} {135301} (\bibinfo {year} {2008})}\BibitemShut {NoStop}%
\bibitem [{\citenamefont {Meppelink}\ \emph {et~al.}(2009)\citenamefont
  {Meppelink}, \citenamefont {Koller},\ and\ \citenamefont {van~der
  Straten}}]{meppelink2009sound}%
  \BibitemOpen
  \bibfield  {author} {\bibinfo {author} {\bibfnamefont {R.}~\bibnamefont
  {Meppelink}}, \bibinfo {author} {\bibfnamefont {S.~B.}\ \bibnamefont
  {Koller}},\ and\ \bibinfo {author} {\bibfnamefont {P.}~\bibnamefont {van~der
  Straten}},\ }\bibfield  {title} {\bibinfo {title} {Sound propagation in a
  {B}ose--{E}instein condensate at finite temperatures},\ }\href
  {https://doi.org/10.1103/PhysRevA.80.043605} {\bibfield  {journal} {\bibinfo
  {journal} {Phys. Rev. A}\ }\textbf {\bibinfo {volume} {80}},\ \bibinfo
  {pages} {043605} (\bibinfo {year} {2009})}\BibitemShut {NoStop}%
\bibitem [{\citenamefont {Jaskula}\ \emph {et~al.}(2012)\citenamefont
  {Jaskula}, \citenamefont {Partridge}, \citenamefont {Bonneau}, \citenamefont
  {Lopes}, \citenamefont {Ruaudel}, \citenamefont {Boiron},\ and\ \citenamefont
  {Westbrook}}]{jaskula2012acoustic}%
  \BibitemOpen
  \bibfield  {author} {\bibinfo {author} {\bibfnamefont {J.-C.}\ \bibnamefont
  {Jaskula}}, \bibinfo {author} {\bibfnamefont {G.~B.}\ \bibnamefont
  {Partridge}}, \bibinfo {author} {\bibfnamefont {M.}~\bibnamefont {Bonneau}},
  \bibinfo {author} {\bibfnamefont {R.}~\bibnamefont {Lopes}}, \bibinfo
  {author} {\bibfnamefont {J.}~\bibnamefont {Ruaudel}}, \bibinfo {author}
  {\bibfnamefont {D.}~\bibnamefont {Boiron}},\ and\ \bibinfo {author}
  {\bibfnamefont {C.~I.}\ \bibnamefont {Westbrook}},\ }\bibfield  {title}
  {\bibinfo {title} {{Acoustic Analog to the Dynamical Casimir Effect in a
  Bose--Einstein Condensate}},\ }\href
  {https://doi.org/10.1103/PhysRevLett.109.220401} {\bibfield  {journal}
  {\bibinfo  {journal} {Phys. Rev. Lett.}\ }\textbf {\bibinfo {volume} {109}},\
  \bibinfo {pages} {220401} (\bibinfo {year} {2012})}\BibitemShut {NoStop}%
\bibitem [{\citenamefont {Lopes}\ \emph {et~al.}(2017)\citenamefont {Lopes},
  \citenamefont {Eigen}, \citenamefont {Barker}, \citenamefont {Viebahn},
  \citenamefont {Robert-de Saint-Vincent}, \citenamefont {Navon}, \citenamefont
  {Hadzibabic},\ and\ \citenamefont {Smith}}]{lopes2017quasiparticle}%
  \BibitemOpen
  \bibfield  {author} {\bibinfo {author} {\bibfnamefont {R.}~\bibnamefont
  {Lopes}}, \bibinfo {author} {\bibfnamefont {C.}~\bibnamefont {Eigen}},
  \bibinfo {author} {\bibfnamefont {A.}~\bibnamefont {Barker}}, \bibinfo
  {author} {\bibfnamefont {K.~G.~H.}\ \bibnamefont {Viebahn}}, \bibinfo
  {author} {\bibfnamefont {M.}~\bibnamefont {Robert-de Saint-Vincent}},
  \bibinfo {author} {\bibfnamefont {N.}~\bibnamefont {Navon}}, \bibinfo
  {author} {\bibfnamefont {Z.}~\bibnamefont {Hadzibabic}},\ and\ \bibinfo
  {author} {\bibfnamefont {R.~P.}\ \bibnamefont {Smith}},\ }\bibfield  {title}
  {\bibinfo {title} {{Quasiparticle Energy in a Strongly Interacting
  Homogeneous {Bose--Einstein} Condensate}},\ }\href
  {https://doi.org/10.1103/PhysRevLett.118.210401} {\bibfield  {journal}
  {\bibinfo  {journal} {Phys. Rev. Lett.}\ }\textbf {\bibinfo {volume} {118}},\
  \bibinfo {pages} {210401} (\bibinfo {year} {2017})}\BibitemShut {NoStop}%
\bibitem [{\citenamefont {Clark}\ \emph {et~al.}(2017)\citenamefont {Clark},
  \citenamefont {Gaj}, \citenamefont {Feng},\ and\ \citenamefont
  {Chin}}]{clark2017collective}%
  \BibitemOpen
  \bibfield  {author} {\bibinfo {author} {\bibfnamefont {L.~W.}\ \bibnamefont
  {Clark}}, \bibinfo {author} {\bibfnamefont {A.}~\bibnamefont {Gaj}}, \bibinfo
  {author} {\bibfnamefont {L.}~\bibnamefont {Feng}},\ and\ \bibinfo {author}
  {\bibfnamefont {C.}~\bibnamefont {Chin}},\ }\bibfield  {title} {\bibinfo
  {title} {{Collective emission of matter-wave jets from driven Bose--Einstein
  condensates}},\ }\href {https://doi.org/10.1038/nature24272} {\bibfield
  {journal} {\bibinfo  {journal} {Nature}\ }\textbf {\bibinfo {volume} {551}},\
  \bibinfo {pages} {356} (\bibinfo {year} {2017})}\BibitemShut {NoStop}%
\bibitem [{\citenamefont {Ville}\ \emph {et~al.}(2018)\citenamefont {Ville},
  \citenamefont {Saint-Jalm}, \citenamefont {Le~Cerf}, \citenamefont
  {Aidelsburger}, \citenamefont {Nascimb\`ene}, \citenamefont {Dalibard},\ and\
  \citenamefont {Beugnon}}]{ville2018sound}%
  \BibitemOpen
  \bibfield  {author} {\bibinfo {author} {\bibfnamefont {J.~L.}\ \bibnamefont
  {Ville}}, \bibinfo {author} {\bibfnamefont {R.}~\bibnamefont {Saint-Jalm}},
  \bibinfo {author} {\bibfnamefont {E.}~\bibnamefont {Le~Cerf}}, \bibinfo
  {author} {\bibfnamefont {M.}~\bibnamefont {Aidelsburger}}, \bibinfo {author}
  {\bibfnamefont {S.}~\bibnamefont {Nascimb\`ene}}, \bibinfo {author}
  {\bibfnamefont {J.}~\bibnamefont {Dalibard}},\ and\ \bibinfo {author}
  {\bibfnamefont {J.}~\bibnamefont {Beugnon}},\ }\bibfield  {title} {\bibinfo
  {title} {{Sound Propagation in a Uniform Superfluid Two-Dimensional Bose
  Gas}},\ }\href {https://doi.org/10.1103/PhysRevLett.121.145301} {\bibfield
  {journal} {\bibinfo  {journal} {Phys. Rev. Lett.}\ }\textbf {\bibinfo
  {volume} {121}},\ \bibinfo {pages} {145301} (\bibinfo {year}
  {2018})}\BibitemShut {NoStop}%
\bibitem [{\citenamefont {Garratt}\ \emph {et~al.}(2019)\citenamefont
  {Garratt}, \citenamefont {Eigen}, \citenamefont {Zhang}, \citenamefont
  {Turz\'ak}, \citenamefont {Lopes}, \citenamefont {Smith}, \citenamefont
  {Hadzibabic},\ and\ \citenamefont {Navon}}]{garratt2019from}%
  \BibitemOpen
  \bibfield  {author} {\bibinfo {author} {\bibfnamefont {S.~J.}\ \bibnamefont
  {Garratt}}, \bibinfo {author} {\bibfnamefont {C.}~\bibnamefont {Eigen}},
  \bibinfo {author} {\bibfnamefont {J.}~\bibnamefont {Zhang}}, \bibinfo
  {author} {\bibfnamefont {P.}~\bibnamefont {Turz\'ak}}, \bibinfo {author}
  {\bibfnamefont {R.}~\bibnamefont {Lopes}}, \bibinfo {author} {\bibfnamefont
  {R.~P.}\ \bibnamefont {Smith}}, \bibinfo {author} {\bibfnamefont
  {Z.}~\bibnamefont {Hadzibabic}},\ and\ \bibinfo {author} {\bibfnamefont
  {N.}~\bibnamefont {Navon}},\ }\bibfield  {title} {\bibinfo {title} {{From
  single-particle excitations to sound waves in a box-trapped atomic
  Bose--Einstein condensate}},\ }\href
  {https://doi.org/10.1103/PhysRevA.99.021601} {\bibfield  {journal} {\bibinfo
  {journal} {Phys. Rev. A}\ }\textbf {\bibinfo {volume} {99}},\ \bibinfo
  {pages} {021601} (\bibinfo {year} {2019})}\BibitemShut {NoStop}%
\bibitem [{\citenamefont {Feng}\ \emph {et~al.}(2019)\citenamefont {Feng},
  \citenamefont {Hu}, \citenamefont {Clark},\ and\ \citenamefont
  {Chin}}]{feng2019correlations}%
  \BibitemOpen
  \bibfield  {author} {\bibinfo {author} {\bibfnamefont {L.}~\bibnamefont
  {Feng}}, \bibinfo {author} {\bibfnamefont {J.}~\bibnamefont {Hu}}, \bibinfo
  {author} {\bibfnamefont {L.~W.}\ \bibnamefont {Clark}},\ and\ \bibinfo
  {author} {\bibfnamefont {C.}~\bibnamefont {Chin}},\ }\bibfield  {title}
  {\bibinfo {title} {Correlations in high-harmonic generation of matter-wave
  jets revealed by pattern recognition},\ }\href
  {https://doi.org/10.1126/science.aat5008} {\bibfield  {journal} {\bibinfo
  {journal} {Science}\ }\textbf {\bibinfo {volume} {363}},\ \bibinfo {pages}
  {521} (\bibinfo {year} {2019})}\BibitemShut {NoStop}%
\bibitem [{\citenamefont {Bogoliubov}(1947)}]{bogoliubov1947on}%
  \BibitemOpen
  \bibfield  {author} {\bibinfo {author} {\bibfnamefont {N.~N.}\ \bibnamefont
  {Bogoliubov}},\ }\bibfield  {title} {\bibinfo {title} {On the theory of
  superfluidity},\ }\href@noop {} {\bibfield  {journal} {\bibinfo  {journal}
  {J. Phys. USSR}\ }\textbf {\bibinfo {volume} {11}},\ \bibinfo {pages} {23}
  (\bibinfo {year} {1947})}\BibitemShut {NoStop}%
\bibitem [{\citenamefont {Beliaev}(1958)}]{beliaev1958energy}%
  \BibitemOpen
  \bibfield  {author} {\bibinfo {author} {\bibfnamefont {S.~T.}\ \bibnamefont
  {Beliaev}},\ }\bibfield  {title} {\bibinfo {title} {{Energy spectrum of a
  non-ideal Bose gas}},\ }\href@noop {} {\bibfield  {journal} {\bibinfo
  {journal} {Sov. Phys. JETP}\ }\textbf {\bibinfo {volume} {34}},\ \bibinfo
  {pages} {299} (\bibinfo {year} {1958})}\BibitemShut {NoStop}%
\bibitem [{\citenamefont {Navon}\ \emph {et~al.}(2016)\citenamefont {Navon},
  \citenamefont {Gaunt}, \citenamefont {Smith},\ and\ \citenamefont
  {Hadzibabic}}]{navon2016emergence}%
  \BibitemOpen
  \bibfield  {author} {\bibinfo {author} {\bibfnamefont {N.}~\bibnamefont
  {Navon}}, \bibinfo {author} {\bibfnamefont {A.~L.}\ \bibnamefont {Gaunt}},
  \bibinfo {author} {\bibfnamefont {R.~P.}\ \bibnamefont {Smith}},\ and\
  \bibinfo {author} {\bibfnamefont {Z.}~\bibnamefont {Hadzibabic}},\ }\bibfield
   {title} {\bibinfo {title} {Emergence of a turbulent cascade in a quantum
  gas},\ }\href {https://doi.org/10.1038/nature20114} {\bibfield  {journal}
  {\bibinfo  {journal} {Nature}\ }\textbf {\bibinfo {volume} {539}},\ \bibinfo
  {pages} {72} (\bibinfo {year} {2016})}\BibitemShut {NoStop}%
\bibitem [{\citenamefont {Gaunt}\ \emph {et~al.}(2013)\citenamefont {Gaunt},
  \citenamefont {Schmidutz}, \citenamefont {Gotlibovych}, \citenamefont
  {Smith},\ and\ \citenamefont {Hadzibabic}}]{gaunt2013bose}%
  \BibitemOpen
  \bibfield  {author} {\bibinfo {author} {\bibfnamefont {A.~L.}\ \bibnamefont
  {Gaunt}}, \bibinfo {author} {\bibfnamefont {T.~F.}\ \bibnamefont
  {Schmidutz}}, \bibinfo {author} {\bibfnamefont {I.}~\bibnamefont
  {Gotlibovych}}, \bibinfo {author} {\bibfnamefont {R.~P.}\ \bibnamefont
  {Smith}},\ and\ \bibinfo {author} {\bibfnamefont {Z.}~\bibnamefont
  {Hadzibabic}},\ }\bibfield  {title} {\bibinfo {title} {{Bose--Einstein
  Condensation of Atoms in a Uniform Potential}},\ }\href
  {https://doi.org/10.1103/PhysRevLett.110.200406} {\bibfield  {journal}
  {\bibinfo  {journal} {Phys. Rev. Lett.}\ }\textbf {\bibinfo {volume} {110}},\
  \bibinfo {pages} {200406} (\bibinfo {year} {2013})}\BibitemShut {NoStop}%
\bibitem [{Sup()}]{Supplementary}%
  \BibitemOpen
  \href@noop {} {}\bibinfo {note} {Supplemental Material.}\BibitemShut {Stop}%
\bibitem [{Note1()}]{Note1}%
  \BibitemOpen
  \bibinfo {note} {More generally, a $M$-body process of this form would lead
  to a $1/(2M-1)$ power-law scaling (see Section~VI in \cite
  {Supplementary}).}\BibitemShut {Stop}%
\bibitem [{nld()}]{nldamp-FN-THG}%
  \BibitemOpen
  \href@noop {} {}\bibinfo {note} {We extract $\mathcal{F}_3$ by fitting the
  data with $v(\ts)= \mathcal{F}_1 \sin(\omega \ts+\phi_1) + \mathcal{F}_3
  \sin(3\omega \ts+\phi_3)$.}\BibitemShut {Stop}%
\bibitem [{Note2()}]{Note2}%
  \BibitemOpen
  \bibinfo {note} {In this case the response amplitude still follows the
  resonant scaling $v_\protect \textrm {m}\propto U_{\protect \textrm
  {s}}^{1/3}$ seen in Fig.~\ref {Fig2}(b).}\BibitemShut {Stop}%
\bibitem [{\citenamefont {Navon}\ \emph {et~al.}(2019)\citenamefont {Navon},
  \citenamefont {Eigen}, \citenamefont {Zhang}, \citenamefont {Lopes},
  \citenamefont {Gaunt}, \citenamefont {Fujimoto}, \citenamefont {Tsubota},
  \citenamefont {Smith},\ and\ \citenamefont
  {Hadzibabic}}]{navon2019synthetic}%
  \BibitemOpen
  \bibfield  {author} {\bibinfo {author} {\bibfnamefont {N.}~\bibnamefont
  {Navon}}, \bibinfo {author} {\bibfnamefont {C.}~\bibnamefont {Eigen}},
  \bibinfo {author} {\bibfnamefont {J.}~\bibnamefont {Zhang}}, \bibinfo
  {author} {\bibfnamefont {R.}~\bibnamefont {Lopes}}, \bibinfo {author}
  {\bibfnamefont {A.~L.}\ \bibnamefont {Gaunt}}, \bibinfo {author}
  {\bibfnamefont {K.}~\bibnamefont {Fujimoto}}, \bibinfo {author}
  {\bibfnamefont {M.}~\bibnamefont {Tsubota}}, \bibinfo {author} {\bibfnamefont
  {R.~P.}\ \bibnamefont {Smith}},\ and\ \bibinfo {author} {\bibfnamefont
  {Z.}~\bibnamefont {Hadzibabic}},\ }\bibfield  {title} {\bibinfo {title}
  {Synthetic dissipation and cascade fluxes in a turbulent quantum gas},\
  }\href {https://doi.org/10.1126/science.aau6103} {\bibfield  {journal}
  {\bibinfo  {journal} {Science}\ }\textbf {\bibinfo {volume} {366}},\ \bibinfo
  {pages} {382} (\bibinfo {year} {2019})}\BibitemShut {NoStop}%
\end{thebibliography}

%

%%%%%%%%%%%%%%%%%%%%%%%%%%%%%%%%
%%%%%%%%%%%%%%%%%%%%%%%%%%%%%%%%

\newpage
\cleardoublepage

\setcounter{figure}{0} 
\setcounter{equation}{0} 

\renewcommand\theequation{S\arabic{equation}} 
\renewcommand\thefigure{S\arabic{figure}} 

\section{Supplemental Material}

%%%%%%%%%%%%%%%%%%%%%%%%%%%%%%%%
%%%%%%%%%%%%%%%%%%%%%%%%%%%%%%%%

\subsection{\textsc{I.~Classical Oscillator Model with Nonlinear Damping}}

The equation of motion of a general classical model for a driven oscillator where the damping is proportional to the $\eta$-th power of the velocity is given by
\begin{equation}
    \ddot{x}+\gamma_{\eta} \dot{x}^{\eta}+\omega_d^2 x=U \sin \omega t    \,,
\end{equation}
which for $\eta>1$ leads to nonlinear dynamics, including the generation of higher-order harmonics. 
However, in cases where the system reaches a steady state oscillating predominantly at $\omega$ (\emph{i.e.}~small higher-harmonic amplitudes) the response amplitude can be estimated from the main Fourier component $\omega$:
\begin{align}
    (\omega_d^2-\omega^2) x_0+q \gamma_{\eta} \omega^{\eta}x^{\eta}_0 = U\, ,
\end{align}
where $q$ is a prefactor that depends on $\eta$.
 In the case of a resonant drive ($\omega_d=\omega$) the amplitude simply scales as $x_0\sim U^{1/{\eta}}$. 
Hence for a scaling $v\propto U^{1/3}$ we identify a cubic damping term with $\eta=3$; note that extending the description to also include position-dependent damping, \emph{i.e.}~a term $\propto x^{\eta-w} \dot{x}^{w}$, where $0\leq w\leq\eta$, does not alter this scaling. 

\subsection{\textsc{II. Derivation of the Reduced Bogoliubov Hamiltonian}}

The Hamiltonian describing weakly interacting bosons in a cylindrical box trap is%
\begin{eqnarray}
\hat{H} &=&\int \text{d}^{3}r\;\hat{\psi}^{\dag }(\mathbf{r})\left\{ -\frac{\hbar ^{2}}{2m}%
\nabla ^{2}-\mu +V(\mathbf{r})\right\} \hat{\psi}(\mathbf{r})  \notag \\
&&+\frac{1}{2}g\int \text{d}^3r\;\hat{\psi}^{\dag }(\mathbf{r})\hat{\psi}^{\dag }(%
\mathbf{r})\hat{\psi}(\mathbf{r})\hat{\psi}(\mathbf{r}),
\end{eqnarray}%
where $\hat{\psi}(\mathbf{r})$ is the boson field operator, $g=
4\pi \hbar ^{2}a/m$ the $s$-wave interaction parameter, and $V(\mathbf{r})$ the box potential %
\begin{eqnarray}
V\left( \rho ,\phi ,z\right)  &=&0,\quad 0<z<L, \,\, 0<\rho <R \notag \\
&=&{\infty, \quad \mbox{otherwise.}}
\end{eqnarray}%
We employ the Bogoliubov approximation, splitting the field
operator into two parts $\hat{\psi}(\mathbf{r})=\varphi (\mathbf{r})+\delta
\hat{\psi}(\mathbf{r})$, where $\varphi (\mathbf{r})=\left\langle \hat{\psi}(%
\mathbf{r})\right\rangle $ is the condensate wave function, and $\delta
\hat{\psi}(\mathbf{r})$ is the fluctuation. We expand the Hamiltonian in the order of fluctuation as $\hat{H}=H^{\left( 0\right) }+\hat{H}^{\left(
1\right) }+\hat{H}^{\left( 2\right) }+\cdots $, where
\begin{eqnarray}
H^{\left( 0\right) } &=&\int \text{d}^3r\;\varphi ^{\ast }(\mathbf{r})\left\{ -%
\frac{\hbar ^{2} \nabla ^{2} }{2m}-\mu \right.  \notag \\
&&+\left. V(\mathbf{r})+\frac{1}{2}g\left\vert \varphi (\mathbf{r}%
)\right\vert ^{2}\right\} \varphi (\mathbf{r})\, , \\
\hat{H}^{\left( 1\right) } &=&\int \text{d}^3r\;\delta \hat{\psi}^{\dag }(\mathbf{r}%
)\left\{ -\frac{\hbar ^{2} \nabla ^{2}}{2m}-\mu \right.  \notag \\
&&+\left. V(\mathbf{r})+g\left\vert \varphi (\mathbf{r})\right\vert
^{2}\right\} \varphi (\mathbf{r})+\text{H.c.}\, , \\
\hat{H}^{\left( 2\right) } &=&\int \text{d}^3r\;\delta \hat{\psi}^{\dag }(\mathbf{r}%
)\left\{ -\frac{\hbar ^{2} \nabla ^{2}}{2m}-\mu \right.  \notag \\
&&+\left. V(\mathbf{r})+2g\left\vert \varphi (\mathbf{r})\right\vert
^{2}\right\} \delta \hat{\psi}(\mathbf{r})  \notag \\
&&+\frac{1}{2}g\int \text{d}^3r\;\left\{ \varphi ^{\ast 2}(\mathbf{r})\delta \hat{%
\psi}^{2}(\mathbf{r})+\text{H.c.}\right\} \,.
\end{eqnarray}%
Here $H^{\left( 0\right) }$ is the mean-field energy, and minimizing $H^{\left( 0\right) }$ leads to the Gross--Pitaevskii equation,
\begin{equation}
\left\{ -\frac{\hbar ^{2} \nabla ^{2} }{2M}+V(\mathbf{r})+g\left\vert \varphi (%
\mathbf{r})\right\vert ^{2}\right\} \varphi (\mathbf{r})=\mu \varphi (%
\mathbf{r}).
\end{equation}%
The resultant ground state condensate wave function $\varphi(\mathbf{r})$ eliminates the first order Hamiltonian $\hat{H}^{(1)}$, and we have
\begin{equation}
\hat{H} = E^{(0)} + \hat{H}^{(2)} + \cdots.
\label{H2}
\end{equation}%
Performing the Bogoliubov transformation
\begin{eqnarray}
\delta \hat{\psi}(\mathbf{r}) &=&\sum\limits_{i }\left[ u_{i }(%
\mathbf{r})\hat{a}_{i }+v_{i }^{\ast }(\mathbf{r})\hat{a}_{i}^{\dag }\right],
\label{BdG_TF01}
\\
\delta \hat{\psi}^{\dag }(\mathbf{r}) &=&\sum\limits_{i }\left[
u_{i }^{\ast }(\mathbf{r})\hat{a}_{i }^{\dag }+v_{i }(%
\mathbf{r})\hat{a}_{i }\right]\,,
\label{BdG_TF02}
\end{eqnarray}%
diagonalizes the Hamiltonian to second order
\begin{equation}
\hat{H}=E_{\rm gs}+\sum\limits_{i }E_{i }\hat{a}_{i }^{\dag }\hat{a}%
_{i } + \cdots \,.
\end{equation}%
Here $E_{\rm gs}$ is the ground-state energy including the mean-field energy and
the Lee-Huang-Yang correction. The excitation spectrum $E_{i }$ and the
corresponding functions $u_{i }(\mathbf{r}),v_{i }(\mathbf{r})$,
can be obtained by solving the  Bogoliubov equations,%
\begin{eqnarray}
\left[ -\frac{\hbar ^{2} \nabla ^{2}}{2m} - \mu  +V(\mathbf{r})+2g|\varphi(\mathbf{r})|^{2}\right]
u_{i }&& \notag \\
+ g\varphi^{2}(\mathbf{r}) v_{i } &=&E_{i
}u_{i }, \\
\left[ -\frac{\hbar ^{2} \nabla ^{2}}{2m} - \mu +V(\mathbf{r})+2g|\varphi(\mathbf{r})|^{2}\right]
v_{i } && \notag \\
+ g\varphi^{*2}(\mathbf{r}) u_{i } &=&-E_{i
}v_{i }.
\end{eqnarray}%
To take into account the coupling between collective modes, we retain the third-order terms in the expansion of the Hamiltonian:
\begin{equation}
\label{eq:H3term}
\hat{H}^{\left( 3\right) }=g\int \text{d}^3r\;\,\varphi^{\ast}(\mathbf{r})\delta
\hat{\psi}^{\dag }(\mathbf{r})\delta \hat{\psi}(\mathbf{r})\delta \hat{\psi}(%
\mathbf{r})+\text{H.c.}\,\,\,.
\end{equation}%
Substituting Eqs. (\ref{BdG_TF01},~\ref{BdG_TF02}) into Eq.~(\ref{eq:H3term}) yields
\begin{eqnarray}
\hat{H}^{\left( 3\right) } &=&g\sum\limits_{i j k
}\int \text{d}^3r\; \varphi^{*}(\mathbf{r})   \notag \\
&&\times \left[ u_{i }^{\ast }(\mathbf{r})\hat{a}_{i }^{\dag
}+v_{i }(\mathbf{r})\hat{a}_{i }\right]\left[ u_{j }(\mathbf{r})\hat{a}_{j }+v_{j }^{\ast }(%
\mathbf{r})\hat{a}_{j }^{\dag }\right] \notag \\
&& \times \left[ u_{k }(\mathbf{r})%
\hat{a}_{k }+v_{k }^{\ast }(\mathbf{r})\hat{a}_{k }^{\dag }%
\right] + \text{H.c.}\,.
\end{eqnarray}%
Motivated by the experiment, we focus on the coupling between the fundamental $\hat{d}\equiv\hat{a}_{\mathrm{fun}}$ and an auxiliary mode $\hat{b}\equiv\hat{a}_{\mathrm{aux}}$. The terms of the form $\hat{b}^{\dagger}\hat{b}^{\dagger}\hat{d}$ lead to a linear damping of the fundamental mode, however the large energy difference highly suppresses this process. The terms of the form $\hat{b}^{\dagger}\hat{d}\hat{d}$ describe the nonlinear decay process of Fig.~\ref{Fig1}(b).
The resulting reduced Hamiltonian of our model in the absence of forcing and dissipation is thus
\begin{eqnarray}
\hat{H}_\textrm{red} &=& \hbar \omega _{d}\hat{d}^{\dag }\hat{d}+\hbar \omega _{b}\hat{b}^{\dag }\hat{b}%
+\hbar \lambda \left( \hat{b}^{\dag }\hat{d}\hat{d}+\text{H.c.}\right), 
\end{eqnarray}%
where the corresponding coupling strength is given by
\begin{eqnarray}
\hbar\lambda &\equiv& g \int \text{d}^3r\; \left\{ \varphi^{*}(\mathbf{r})
\left( 
u_{\mathrm{aux}}^{*}(\mathbf{r}) u_{\mathrm{fun}}(\mathbf{r})
+ 2 v_{\mathrm{aux}}^{*}(\mathbf{r}) v_{\mathrm{fun}}(\mathbf{r})
\right)
u_{\mathrm{fun}}(\mathbf{r}) \right. \notag
\\
&& + \left. \varphi (\mathbf{r})
\left( 
v_{\mathrm{aux}}^{*}(\mathbf{r}) v_{\mathrm{fun}}(\mathbf{r})
+ 2 u_{\mathrm{aux}}^{*}(\mathbf{r}) u_{\mathrm{fun}}(\mathbf{r})
\right)
v_{\mathrm{fun}}(\mathbf{r})\right\}\,.
\end{eqnarray}%

\subsection{\textsc{III.~Effective Equation of Motion for the Fundamental Mode}}

In this section we introduce a closed effective equation of motion for the fundamental mode, derived from the equations of motion of the coupled fundamental and auxiliary modes.

We include the driving and dissipation in the description of the dynamics of the fundamental and auxiliary modes using the quantum master equation $\hbar\partial _{t}\rho =-i\left[ \hat{H},\rho %
\right] +D\left[ \rho \right] $, where the Hamiltonian $\hat{H}$ and the dissipator $D$ read%
\begin{eqnarray}
\hat{H} &=& \hat{H}_\textrm{red} + \hbar\Omega \sin \left( \omega t \right) \left( \hat{d}^{\dag }+\hat{d}\right) ,
\\
D[\rho] &=& \hbar \gamma _{1}\left( 2\hat{b}\rho \hat{b}^{\dag }-\hat{%
b}^{\dag }\hat{b}\rho -\rho \hat{b}^{\dag }\hat{b}\right) .
\end{eqnarray}%

The auxiliary mode readily couples to other modes, which effectively act as a quasi-continuum to which it can decay.
Here we assume the simplest form of dissipation for the auxiliary mode, a one-particle loss process with loss rate $\gamma_{1}$. 
In terms of the experimental  parameters, the scaled drive amplitude is
\begin{equation}
\Omega = \alpha \Us/(\hbar L) \, ,
\end{equation}
where 
\begin{equation}
\alpha = \int \text{d}^3r\; \left(
v_{\mathrm{fun}}(\mathbf{r}) z \varphi(\mathbf{r}) + u_{\mathrm{fun}}(\mathbf{r}) z \varphi^{*}(\mathbf{r})\right) \, .
\end{equation}
For a cylindrical-box trapped BEC in the Thomas--Fermi regime we have
\begin{equation}
\alpha \approx L \sqrt{N} \frac{2^{5/4}}{\pi^{3/2}}\sqrt{\frac{\xi}{L}} \, .
\end{equation}

Using a functional integral method, one can derive the following effective equation of motion for the reduced density matrix of the fundamental mode $\rho_{d}$:
\begin{equation}
\label{eq:effmastereq1}
\hbar \partial _{t}\rho_{d} =-i\left[\hat{H}_{\mathrm{eff}},\rho_{d} \right] +D_{\mathrm{eff}}\left[ \rho_{d} \right]\,,
\end{equation}%
where the effective Hamiltonian $\hat{H}_{\mathrm{eff}}$ and dissipator $D_{\mathrm{eff}}$ are given by%
\begin{eqnarray}
\label{eq:effH}
\hat{H}_{\mathrm{eff}} &=&\hbar\omega _{d}\hat{d}^{\dag }\hat{d} + \tfrac{\hbar}{2}\mathrm{Re}[\kappa_{2}]\hat{d}^{\dag }\hat{d}%
^{\dag }\hat{d}\hat{d}+ \notag \\
&& \hbar \Omega \sin \left( \omega t \right) \left( \hat{d}%
^{\dag }+\hat{d}\right), \\
\label{eq:effD}
D_{\mathrm{eff}}[\rho_{d}] &=&-\tfrac{\hbar}{2}\mathrm{Im}[\kappa _{2}] \left( 2\hat{d}\hat{d}\rho_{d} \hat{d}%
^{\dag }\hat{d}^{\dag } \right. \notag \\
&&\left.-\hat{d}^{\dag }\hat{d}^{\dag }\hat{d}\hat{d}\rho_{d}
-\rho_{d} \hat{d}^{\dag }\hat{d}^{\dag }\hat{d}\hat{d}\right).
\end{eqnarray}%
Here the effective nonlinear parameter $\kappa_{2}$ is
\begin{equation}
\label{eq:kappa2micro}
    \kappa _{2} =\frac{2\lambda^{2}}
    { 2\omega_{d} - \omega_{b} + i\gamma _{1}},
\end{equation}
which arises from the perturbative elimination of the auxiliary mode, under the assumption that its occupation remains small.
The coupling to the auxiliary mode not only gives the
fundamental mode an effective self-interaction, given by $\mathrm{Re}[\kappa_{2}]$, but also leads to its
effective nonlinear dissipation, given by $\mathrm{Im}[\kappa_{2}]$.
Note that Eqs.~(\ref{eq:effmastereq1}-\ref{eq:effD}) are robust both with respect to the form of dissipation of the auxiliary mode, and the number of auxiliary excitations involved in the elementary interaction process; only the microscopic details [the form of Eq.~(\ref{eq:kappa2micro})] would change.

\subsection{\textsc{IV.~Mean-Field Analysis of the Equation of Motion for the Fundamental Mode}}

By employing a mean-field approximation in Eq.~(\ref{eq:effmastereq1}) we obtain the equation of motion of the mean dipole mode $d(t) \equiv \langle \hat{d}(t) \rangle$ :
\begin{equation}
i\partial _{t}d=\left( \omega _{d}+\kappa_{2}|d|^{2} \right) d+ \Omega \sin \left( \omega t \right).
\label{eq:eqofm}
\end{equation}

In the limit of weak drive, one can use a harmonic ansatz $d\left( t\right) =A e^{-i\omega t} = - |A| e^{-i(\omega t +\phi)}$ in Eq.~(\ref{eq:eqofm}) and within a rotating wave approximation obtain
\begin{equation}
\left( \omega -\omega _{d}- \kappa_{2}\left\vert
A\right\vert ^{2}\right) \left\vert A\right\vert =- i\frac{\Omega}{2} e^{i \phi }\,,
\end{equation}%
such that
\begin{eqnarray}
\frac{\Omega ^{2}}{4} &=&\left[ \left( \omega -\omega _{d}- \mathrm{Re}[\kappa_{2}] \left\vert
A\right\vert ^{2}\right) ^{2} \right. \notag \\
&& \left.+\left( \mathrm{Im}[\kappa_{2}] \left\vert A\right\vert
^{2}\right) ^{2}\right] \left\vert A\right\vert ^{2}, \\
\tan \left( \phi \right) &=&\frac{\omega -\omega
_{d}-\mathrm{Re}[\kappa_{2}]\left\vert A\right\vert ^{2}}{\mathrm{Im}[\kappa_{2}]\left\vert A\right\vert
^{2}}\,.
\end{eqnarray}%

For $\omega =\omega _{d}$,
\begin{eqnarray}
\left\vert A\right\vert &=&\left( \frac{\Omega}{2|\kappa_{2}|}%
\right) ^{1/3}, \\
\tan \left( \phi \right) &=&-\frac{\mathrm{Re}[\kappa_{2}]}{\mathrm{Im}[\kappa_{2}]}\,,
\end{eqnarray}%
revealing that on resonance $|A| \propto \Omega ^{1/3}$, while the phase shift of the oscillation with respect to the drive reflects the phase of the complex nonlinear parameter $\kappa_2$.

Finally, we relate $d(t)$ to our experimental observable, the mean center-of-mass velocity $v(t)$.
The center-of-mass position operator is
\begin{equation}
    \hat{z}=\frac{1}{N}\int \text{d}^3r\; \hat{\psi}^{\dag }(\mathbf{r})z \hat{\psi}(\mathbf{r}).
\end{equation}
Assuming that the dominant contribution to $\hat{z}$ is due to the fundamental mode, we have $\hat{z}\approx \frac{\alpha}{N} (\hat{d}+\hat{d}^\dag)$.
 Within the mean-field approximation one thus finds that the mean center-of-mass velocity is
\begin{equation}
    v(t)=\langle \partial_t\hat{z}\rangle \approx\frac{2\alpha}{N} \partial_t\text{Re}[d(t)].    
\end{equation}

\subsection{\textsc{V.~Third-Harmonic Generation}}
Intriguingly, Eq.~(\ref{eq:eqofm}) [Eq.~(\ref{eq:dipole}) in the main text] includes the possibility of third-harmonic generation for our system parameters.
To study this effect, we use an ansatz of the form
\begin{equation}
d(t) =A_{1}^{+}e^{i \omega t }+A_{1}^{-}e^{-i \omega t}+A_{3}^{-}e^{-i 3\omega t}\,,
\end{equation}
in Eq.~(\ref{eq:eqofm}), and by balancing the harmonics obtain
\begin{widetext}
\begin{eqnarray}
\left[ \omega _{d}-\omega + \kappa_{2}\left(
\left\vert A_{1}^{-}\right\vert ^{2}+2\left\vert A_{1}^{+}\right\vert
^{2}+2\left\vert A_{3}^{-}\right\vert ^{2}\right) \right] A_{1}^{-} 
&=&-i\frac{1}{2}\Omega - 2\kappa_{2}
A_{3}^{-}A_{1}^{+}A_{1}^{-\ast }, \\
\left[ \omega _{d}+\omega + \kappa_{2} \left(
\left\vert A_{1}^{+}\right\vert ^{2}+2\left\vert A_{1}^{-}\right\vert
^{2}+2\left\vert A_{3}^{-}\right\vert ^{2}\right) \right] A_{1}^{+} 
&=&i\frac{1}{2}\Omega - \kappa_{2} A_{3}^{-\ast 
}A_{1}^{-}A_{1}^{-}, \\
\left[ \omega _{d}-3\omega + \kappa_{2} \left(
\left\vert A_{3}^{-}\right\vert ^{2}+2\left\vert A_{1}^{-}\right\vert
^{2}+2\left\vert A_{1}^{+}\right\vert ^{2}\right) \right] A_{3}^{-} 
&=&-\kappa_{2} A_{1}^{+\ast}A_{1}^{-}A_{1}^{-}\,.
\label{A3_EQ}
\end{eqnarray}
\end{widetext}

Note that $A_{3}^{-}$ is excited through the $A_1^{\pm}$ oscillations. Assuming that the former is much smaller than the latter and that $\omega$ is far from resonance, the amplitudes $A_1^\pm$ can be expanded to linear order in $\Omega$ for weak drives:
\begin{eqnarray}
A_{1}^{\pm} &\simeq &\pm \frac{i \Omega}{2\left( \omega _{d}\pm\omega \right) }\,. 
\end{eqnarray}%
Substituting this result into Eq.~(\ref{A3_EQ}), and assuming that $|\kappa_{2}| \left( \left\vert A_{3}^{-}\right\vert ^{2} + 2\left\vert A_{1}^{-}\right\vert
^{2}+2\left\vert A_{1}^{+}\right\vert ^{2}\right) \ll \left\vert \omega
_{d}-3\omega \right\vert $ yields
\begin{equation}
\label{eq:a3eq}
A_{3}^{-}\simeq i\frac{\kappa_{2}}{3\omega-\omega _{d}} \frac{\Omega^3}{8(\omega+\omega_d)(\omega-\omega_d)^2}\,.
\end{equation}%
This reveals that the third-harmonic amplitude is $\left\vert A_{3}^{-}\right\vert \varpropto \Omega^{3}$, in agreement with our measurements shown in Fig.~\ref{Fig4}.
Since only two fundamental phonons couple to the relay state $b$, this process is rather unusual: two (rather than three) off-resonant excitations of energy  $\hbar\omega\approx\hbar\omega_d/3$ (a process $\propto \Omega^2$) convert into a real $d$ excitation of energy $3\hbar\omega\approx\hbar\omega_d$; to conserve energy, a `counter-rotating' phonon of energy $\approx -\hbar\omega_d/3$ is produced by stimulated emission into the driving field, a process $\propto \Omega$.
On the other hand, on the third-harmonic resonance ($3\omega = \omega_d$), the amplitude $|A_3^-|$ is suppressed not by the detuning $\omega_d - 3\omega$ [see Eq.~(\ref{eq:a3eq})] but by the nonlinear damping rate $\propto \kappa_2 \Omega^2$. In this case the third-harmonic amplitude response is $|A_3^-|\propto \Omega^3/\Omega^2$, \emph{i.e.} linear in $\Omega$.

\subsection{\textsc{VI.~Generalization to $M$-Excitation Damping}}
In this section we generalize our model to a damping mechanism mediated by $M$-excitation interactions and show that the on-resonance amplitude is proportional to the $1/(2M-1)$ power of the driving force.

The corresponding effective Hamiltonian and dissipator read%
\begin{eqnarray}
\hat{H}_{\mathrm{eff}} &=&\hbar \omega _{d}\hat{d}^{\dag }\hat{d}+\tfrac{\hbar}{M}\mathrm{Re}[\kappa_{M}](\hat{d}^{\dag})^M (\hat{d})^M\notag \\
&&+\hbar \Omega\sin \left( \omega t \right) \left( \hat{d}^{\dag }+\hat{d}\right) \, ,
\\
D_{\mathrm{eff}}[\rho] &=&-\tfrac{\hbar}{M}\mathrm{Im}[\kappa _{M}] \left( 2(\hat{d})^M \rho
(\hat{d}^{\dag })^M\right. \notag \\
&&\left. -(\hat{d}^{\dag })^M (\hat{d})^M\rho
-\rho (\hat{d}^{\dag})^M (\hat{d})^M\right)\,.
\end{eqnarray}%
The mean-field equation of-motion is%
\begin{equation}
i\partial _{t}d=\left(\omega _{d}+ \kappa_{M}
\left\vert d\right\vert ^{2M-2}\right) d+ \Omega \sin \left( \omega t \right)\,.
\end{equation}%
Using the ansatz $d\left( t\right) =A e^{-i\omega t} = - |A| e^{-i(\omega t +\phi)}$ and balancing the harmonics yields
\begin{equation}
\left( \omega -\omega _{d}- \kappa_{M}\left\vert
A\right\vert ^{2M-2}\right) \left\vert A\right\vert =-i\frac{\Omega}{2} e^{i \phi }\,.
\end{equation}%
On resonance ($\omega=\omega_d$) we have
\begin{equation}
\left\vert A\right\vert =\left( \frac{\Omega}{2|\kappa_{M}|}%
\right) ^{1/\left( 2M-1\right) } \, ,
\end{equation}
and we also identify the resonance width $\Gamma\propto\mathrm{Im}[\kappa_{M}]\left\vert A\right\vert^{2M-2}\propto \Omega^{(2M-2)/(2M-1)}$.

\end{document}